\documentclass[aps,prl,twocolumn,showpacs,superscriptaddress,notitlepage]{revtex4-1}
\usepackage{amsmath,amssymb,graphicx,color}
\usepackage{amsthm}
\usepackage{times}
\usepackage{latexsym}
\usepackage{stmaryrd}
\usepackage[table]{xcolor}
\usepackage{amsfonts}
\usepackage{bm}
\usepackage{bbm}
\usepackage{stmaryrd}
\usepackage{subfigure}
\usepackage{natbib}
\usepackage{appendix}
\usepackage[colorlinks=true,urlcolor=blue,citecolor=blue]{hyperref}
\usepackage[utf8]{inputenc}

\newcommand\braket[1]{\left\langle\textstyle{#1}\right\rangle}

\newcommand\textF{\textrm{F}}
\newcommand\textMF{\textrm{MF}}

\begin{document}
\title{Frustrated Superradiant Phase Transition}
\author{Jinchen Zhao}
\affiliation{Division of Natural and Applied Sciences, Duke Kunshan University, Kunshan, Jiangsu, 215300 China}
\author{Myung-Joong Hwang}
\email{myungjoong.hwang@duke.edu}
\affiliation{Division of Natural and Applied Sciences, Duke Kunshan University, Kunshan, Jiangsu, 215300 China}
\affiliation{Zu Chongzhi Center for Mathematics and Computational Science, Duke Kunshan University, Kunshan, Jiangsu, 215300 China}

\begin{abstract}
Frustration occurs when a system cannot find a lowest-energy configuration due to conflicting constraints. We show that a frustrated superradiant phase transition occurs when the ground-state superradiance of cavity fields due to local light-matter interactions cannot simultaneously minimize the positive photon hopping energies. We solve the Dicke trimer model on a triangle motif with both negative and positive hopping energies and show that the latter results in a six-fold degenerate ground-state manifold in which the translational symmetry is spontaneously broken. In the frustrated superradiant phase, we find that two sets of diverging time and fluctuation scales coexist, one governed by the mean-field critical exponent and another by a novel critical exponent. The latter is associated with the fluctuation in the difference of local order parameters and gives rise to site-dependent photon number critical exponents, which may serve as an experimental probe for the frustrated superradiant phase. We provide a qualitative explanation for the emergence of unconventional critical scalings and demonstrate that they are generic properties of the frustrated superradiant phase at the hand of a one-dimensional Dicke lattice with an odd number of sites. The mechanism for the frustrated superradiant phase transition discovered here applies to any lattice geometries where the anti-ferromagnetic ordering of neighboring sites are incompatible and therefore our work paves the way towards the exploration of frustrated phases of coupled light and matter. 
\end{abstract}
\maketitle

{\it Introduction.---} 
Competing interactions that cannot be simultaneously satisfied induce frustration that is often responsible for exotic physical properties~\cite{Moessner:2006ew,Han.2008,Bramwell.2001}. For interacting spin systems, frustration may prevent symmetry breaking and magnetic order, leading to novel phases characterized by highly degenerate and entangled ground states such as quantum spin liquids~\cite{Savary.2016, Broholm.2020,Bramwell.2001,Balents.2010,Wannier.1950,Anderson.1973,Ortiz-Ambriz.2019}. Controlled quantum systems such as trapped-ions and ultracold atoms have emerged as promising platforms for a quantum simulation of frustrated spin physics thanks to the possibility to engineer desired interactions and lattice geometries~\cite{Kim.2010,Islam:2013ed,Gross.2017,Struck.2011}. 

A superradiant phase transition (SPT) occurs in a system of coupled bosons and spins when the spin-boson coupling strength exceeds a critical value, whereby the bosonic modes exhibit a ground-state superradiance. The Dicke model is a paradigmatic model exhibiting the SPT~\cite{Hepp:1973jt,Wang:1973ky,Emary:2003da,emary_chaos_2003,Dimer:2007da,Nagy.2010,Nataf:2010dy,Viehmann:2011dy,Torre.2013tfj,Kirton.2019}, which describes a single bosonic mode coupled to $N_a$ two-level atoms and has been realized in disparate physical systems including cavity QED systems~~\cite{Baumann:2010js,Baumann:2011io,Klinder:2015df,Zhang.2021wqi,Kroeze.2018} and trapped-ions~\cite{Safavi-Naini.2018,Gilmore.2021}. The SPTs have also been found in various generalizations of the Dicke models including the finite-component systems~\cite{Hwang:2015eq,Hwang:2016cb, Puebla:2017gq, Felicetti.2020,Zhang.2021,Cai.2021,Chen.2021} and lattice systems~\cite{Zou:2014cy}. 

In this paper, we consider the Dicke trimer model where each site on a triangle motif realizes the Dicke model and neighboring sites are coupled by hopping interaction between oscillators. We show that the Dicke trimer with positive hopping energy undergoes a \emph{frustrated} SPT. The mechanism for the frustrated SPT identified here is that the positive hopping energy favors opposite signs for the spontaneous ground-state superradiance of neighboring cavities, a configuration that becomes incompatible with a triangle motif. Therefore, the frustrated SPT can occur in any lattice geometries where the anti-ferromagnetic ordering of neighboring sites is incompatible. We demonstrate this possibility using the $1$D Dicke lattice model with an odd number of lattice sites as an example, for which a negative photon hopping energy leads to a standard SPT~\cite{Zou:2013iu}. Note that previous studies have focused on spin frustrations induced by a multimode cavity with disorder~\cite{Gopalakrishnan.2009,Strack:2011hv,Buchhold.2013,Gopalakrishnan.2012,Marsh.2021,Chiocchetta.2021,Kelly.2021}.

\begin{figure}[b]
\centering
\raisebox{0.1\height}{\includegraphics[width=0.4\linewidth]{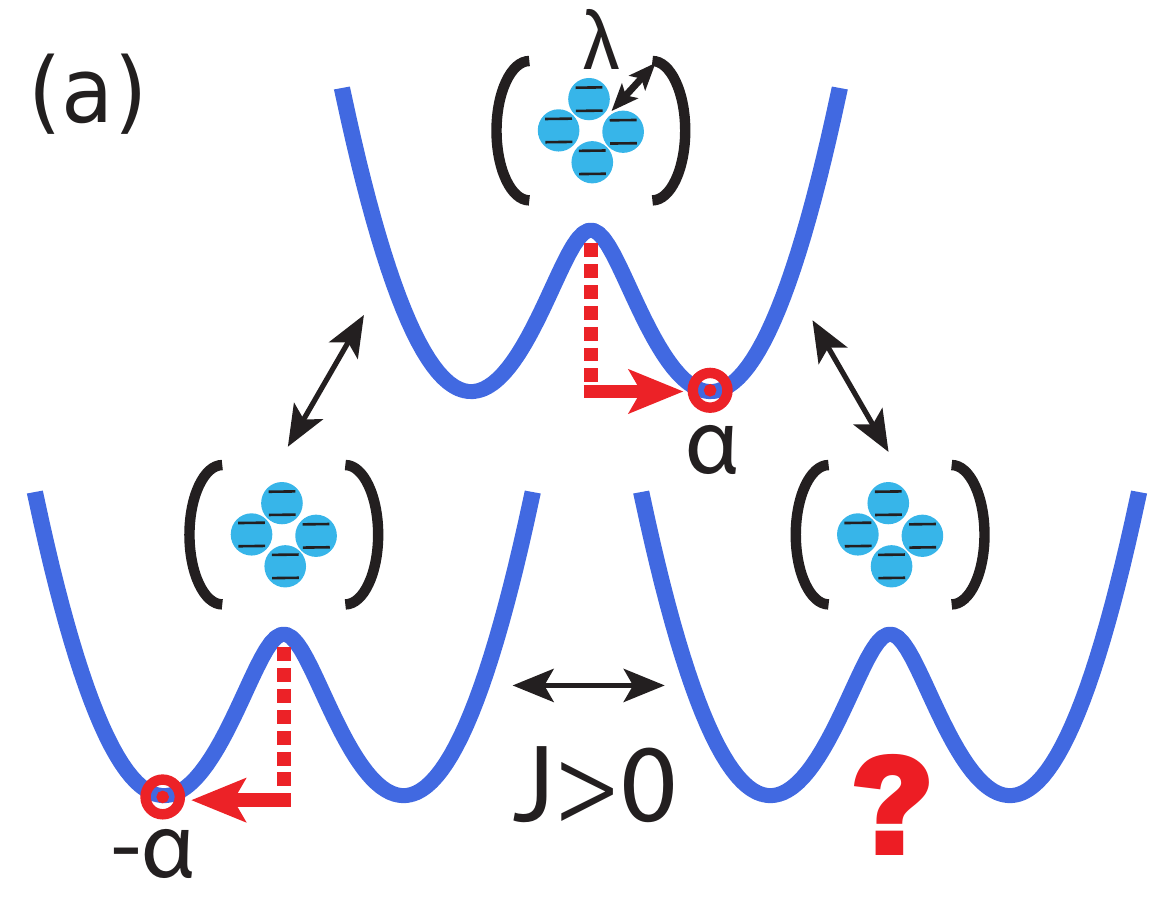}}
\includegraphics[width=0.55\linewidth]{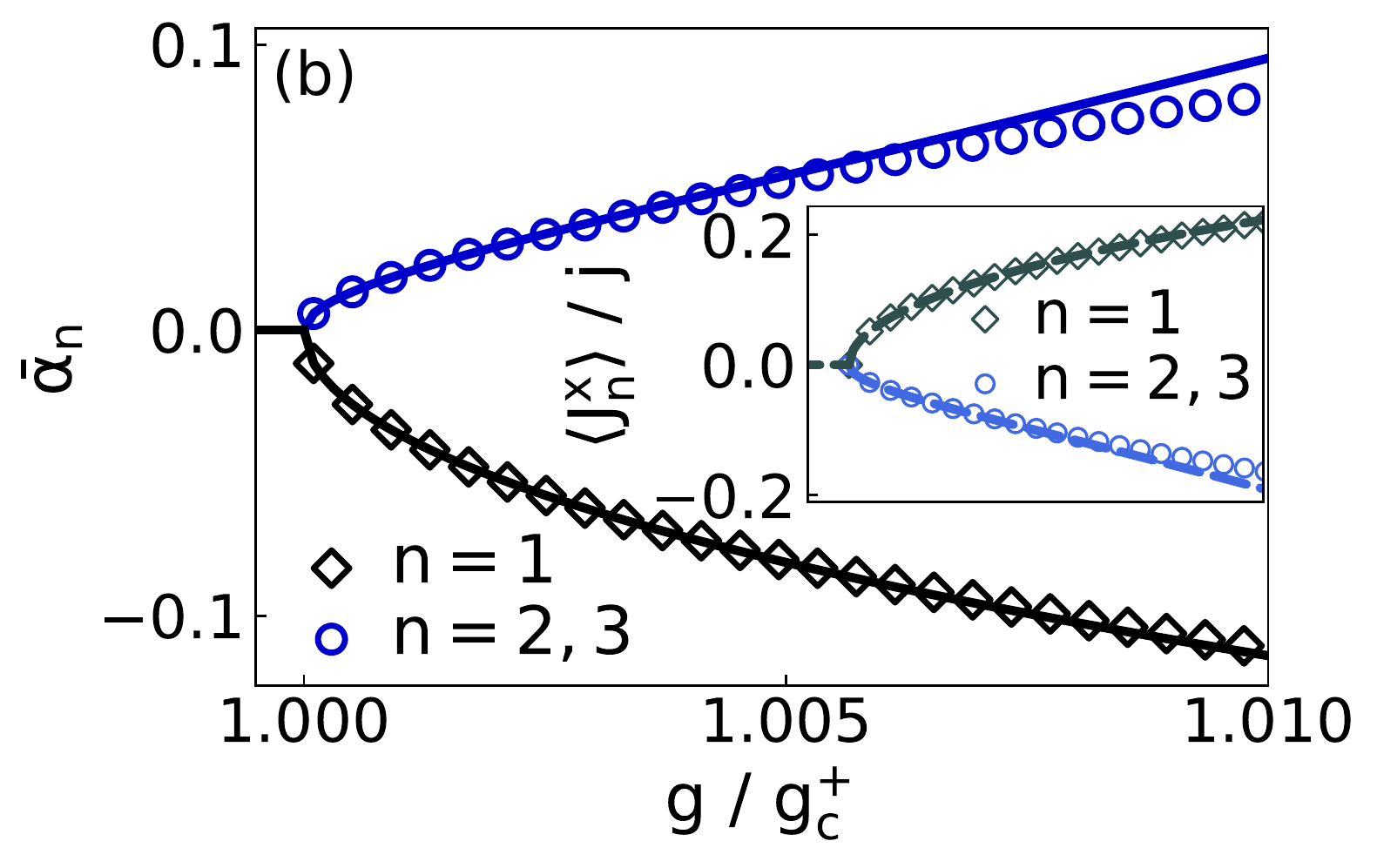} \\
\caption{Frustrated superradiance in the Dicke trimer. (a) The opposite signs for ground-state superradiance $\alpha$ for neighboring cavities, favored by the positive hopping energy $J$, cannot be satisfied on a triangle, leading to geometric frustration analogous to antiferromagnetic triangular Ising spins. (b) One of the six degenerate ground-state solutions for the renormalized cavity and atomic coherence of $n$-th cavity, $\bar{\alpha}_n$ and $\braket{J^x_n}/j$, respectively. Shapes are numerical solutions and lines are approximate analytic solutions. We set $\omega_0=\Omega=1$ and $J=0.01$.}
\label{fig:1}
\end{figure}

In the frustrated superradiant phase (FSP), we find a frustrated ground-state manifold with an extensive degeneracy and a broken translational symmetry, in addition to the broken parity symmetry of the total number of excitations. Moreover, there are two non-degenerate excitations closing their gaps at the critical point with distinct critical exponents: one is the mean-field exponent of the Dicke model~\cite{Emary:2003da} and the other is a novel frustration critical exponent that depends on the lattice size. In addition, we show that the inhomogeneous spatial distribution of the critical excitations results in site-dependent local photon number exponents, which can be used as a probe for the frustrated SPT.

{\it Model.---} We consider a Dicke lattice model, which consists of $N$ coupled cavities, each containing $N_a$ two-level systems realizing the Dicke model, as shown in Fig.~\ref{fig:1}(a) for $N=3$. The Hamiltonian reads
\begin{eqnarray}
\label{eq:hamiltonian}
  H_N = \sum_{n=1}^N \left[H^\textrm{Dicke}_{n}+ J(a_{n}^{\dagger}a_{n+1}+h.c.)\right]
\end{eqnarray}
with  $H^\textrm{Dicke}_n=\omega_{0} a_{n}^{\dagger} a_{n}+\Omega J_{n}^{z}+\frac{2\lambda}{\sqrt{N_a}} \left(a_{n}+a_{n}^{\dagger}\right)J_{n}^{x}$ and $a_{N+1}=a_{1}$. Here, $a_{n}$ is the annihilation operator for the $n$-th cavity with frequency $\omega_0$ and $J_n^{x,y,z}$ are collective spin operators with a total spin $j=N_a/2$ for the $n$-th atomic ensemble with frequency $\Omega$. The on-site atom-photon coupling is $\lambda$ and the photon hopping energy is $J$. The Hamiltonian possesses a global $Z_2$ symmetry as  $[H_N, e^{i \pi \sum_{n=1}^N(a_n^{\dagger} a_n+J_n^{z}+j)}]=0$. We define the mean-value of cavity and atomic coherence for the ground state, $\alpha_n\equiv\braket{a_n}$ and $\left(\braket{J_{n}^{x}}, \braket{J_{n}^{y}}, \braket{J_{n}^{z}}\right)=\frac{N_a}{2}(\sin \theta_n \cos \phi_n, \sin \theta_n \sin \phi_n, \cos \theta_n)$ with $\theta_n \in[0,\pi]$ and $\phi_n \in[0,2 \pi)$.

Let us consider the thermodynamic limit $N_a\rightarrow\infty$. It is convenient to separate the mean-values and fluctuations by $\tilde H_N=U^\dagger H_N U$ with $U=\prod_{n=1}^{N}\mathrm{e}^{-i \phi_n J_{n}^z}\mathrm{e}^{-i \theta_n J_n^{y}}\mathrm{e}^{\alpha_n a_{n}^{\dagger}-\alpha_n^*a_{n}}$ and then apply the Holstein-Primakoff transformation to rotated collective spins, $J^{+}_n=\sqrt{N_a-b_n^{\dagger} b_n}b_n$ and $J_n^{z}=\frac{N_a}{2}-b_n^{\dagger} b_n$, where $b_n$ satisfies $\left[b_n, b_n^{\dagger}\right]=1$~\cite{Emary:2003da,Soldati.2021ex8}. The transformed Hamiltonian is given by~\cite{sup}
\begin{equation}\label{eq:transformedH}
	\tilde{H}_N= E_{\mathrm{GS}}(\{\alpha_n,\theta_n,\phi_n\})+\sum_{s=l,q}H_{s}(\{\alpha_n,\theta_n,\phi_n,a_n,b_n\}),
\end{equation}
where $E_{\mathrm{GS}}$ is the ground-state energy, and $H_q$ ($H_l$) is a quadratic (linear) Hamiltonian in $a_n$ and $b_n$. A SPT occurs when $E_{\mathrm{GS}}$ minimizes for $\alpha_n\neq0$ or $\theta_n\neq\pi$, giving rise to non-zero spontaneous superradiance. We show that the global minimum of $E_{\mathrm{GS}}$ occurs always for $\alpha_n\in \mathbb{R}$ and for \begin{equation}\label{eq:trig}
	\cos{\phi_n}=-\bar\alpha_n/|\bar\alpha_n|,\quad\cos{\theta_n}=-{1}/{\sqrt{1+4g^2\bar{\alpha}_n^2}},
\end{equation}
which leads to
\begin{equation}\label{eq:sqrtE}
	\bar E_\mathrm{GS}(\{\bar\alpha_n\})= \sum_{n=1}^N\left(\bar{\alpha}_n^2 - \frac{1}{2}{\sqrt{1+4g^2\bar{\alpha}_n^2}}+2\bar{J}\bar{\alpha}_{n} \bar{\alpha}_{n+1}\right),
\end{equation}
where $\bar{\alpha}_n=\sqrt{\omega_0/N_a \Omega}~\alpha_n$, $g=2\lambda/\sqrt{\omega_0\Omega}$, $\bar E_{\mathrm{GS}}=E_\mathrm{GS}/(N_a\Omega)$, and $\bar{J}=J/\omega_0$~\cite{sup}.

{\it Frustrated superradiance.---} Let us first introduce the notion of frustrated superradiance using  Eq.~(\ref{eq:sqrtE}). For $J=0$, each cavity takes the form of a double well potential for $g>1$ and $\bar\alpha_n$ has two possible solutions $\bar\alpha_n=\pm\sqrt{g^2-g^{-2}}/2$. A non-zero photon hopping, on the other hand, makes neighboring cavities to be correlated. For $J>0$, opposite signs (anti-aligned) for $\bar\alpha_n$ of the neighboring sites are favored. Therefore, if the underlying lattice geometry is not compatible with anti-aligned configurations for spontaneous superradiance, the system becomes frustrated [Fig.~\ref{fig:1} (a)]. It is the same principle as frustrated Ising spins~\cite{Balents.2010} and here the sign of $\bar\alpha_n$ plays the role of spin direction. We emphasize however that $\bar\alpha_n$ is allowed to vary its magnitude unlike the spin and we will show its striking consequences on the fluctuation below. In contrast, for $J<0$, all sites can be aligned to realize the global minimum without frustration. 

To corroborate this expectation, we first consider the Dicke trimer ($N=3$). For $\bar E_\mathrm{GS}(\{\bar\alpha_n\})$ in Eq. (\ref{eq:sqrtE}), the origin ($\alpha_n=0$ for $\forall n$) is always an extreme point, at which all eigenvalues of Hessian matrix are positive when $g<\operatorname{min}\{\sqrt{1-\bar{J}},\sqrt{1+2\bar{J}}\}$. Therefore, the critical point depends on the sign of $J$,
\begin{equation}
	g_c^{-}=\sqrt{1+2\bar J}~(\textrm{for}~J<0),\quad g_c^{+}=\sqrt{1-\bar J}~(\textrm{for}~J>0),
\end{equation}
where the superscript $\pm$ indicates the sign of $J$. For $g<g_c^{\pm}$, we find the normal phase solution, i.e., $\bar\alpha_n=0$, $\theta_n=\pi$ for $\forall n$ with arbitrary $\phi_n$, which is stable for small hopping rates $-1/2<\bar J<1$. For $J<0$ and $g>g_c^-$, we show that $\bar{E}_\textrm{GS}$ is minimized when $\bar{\alpha}_1=\bar{\alpha}_2=\bar{\alpha}_3$ using Cauchy-Schwarz inequality~\cite{sup}, leading to $\Bar{E}_{\mathrm{GS}}= 3(1+2\bar{J})\Bar{\alpha}^2 - \frac{3}{2} \sqrt{1+4g^2\Bar{\alpha}^2}$. Therefore, we find a non-frustrated superradiant phase (NFSP) solution, $\bar{\alpha}_n=\pm\frac{1}{2g}\sqrt{(g/g_c^-)^4-1}$, which is identical to the single Dicke model except the critical point being shifted by the photon hopping. 

For $J>0$ and $g>g_c^+$, the energy minimum of Eq. (\ref{eq:sqrtE}) cannot be analytically solved. To gain insights beyond numerical solutions shown in Fig.~\ref{fig:1}, we derive an approximate solution for $|g-g_c^+|\ll1$, which reads
\begin{subequations}\label{eq:minsp2a}
	\begin{align}
	\bar{\alpha}_{1}(g) & \simeq-\frac{2 |g-g_c^+|^{1/2}}{\sqrt{3} (g_c^+)^{3/2}} - \frac{|g-g_c^+|^{3/2}}{6 \sqrt{3} (g_c^+)^{5/2}},\\
		\bar{\alpha}_{2}(g)= \bar{\alpha}_{3}(g) & \simeq \frac{|g-g_c^+|^{1/2}}{\sqrt{3} (g_c^+)^{3/2}} +\frac{(8-7\bar{J}) |g-g_c^+|^{3/2}}{12\sqrt{3}\bar{J} (g_c^+)^{5/2}}.
	\end{align}
\end{subequations}
They agree well with the numerical solution for $|g-g_c^+|\ll1$ [Fig.~\ref{fig:1} (b)]. Note that this is one of the six degenerate solutions for which there are two sites having the same sign and a remaining site with the opposite sign, analogously to the antiferromagnetic triangular Ising spins~\cite{Balents.2010}. We call the former a \emph{ferromagnetic} pair and the latter an \emph{unpaired} site. Unlike the spin however the amplitude is not uniform: the ferromagnetic pair has the same magnitude, which is different from that of the unpaired site. In fact, we prove that this amplitude relation holds for any values of $g>g_c^+$~\cite{sup}.

\begin{figure}[t]
\centering
\includegraphics[width=0.5\linewidth]{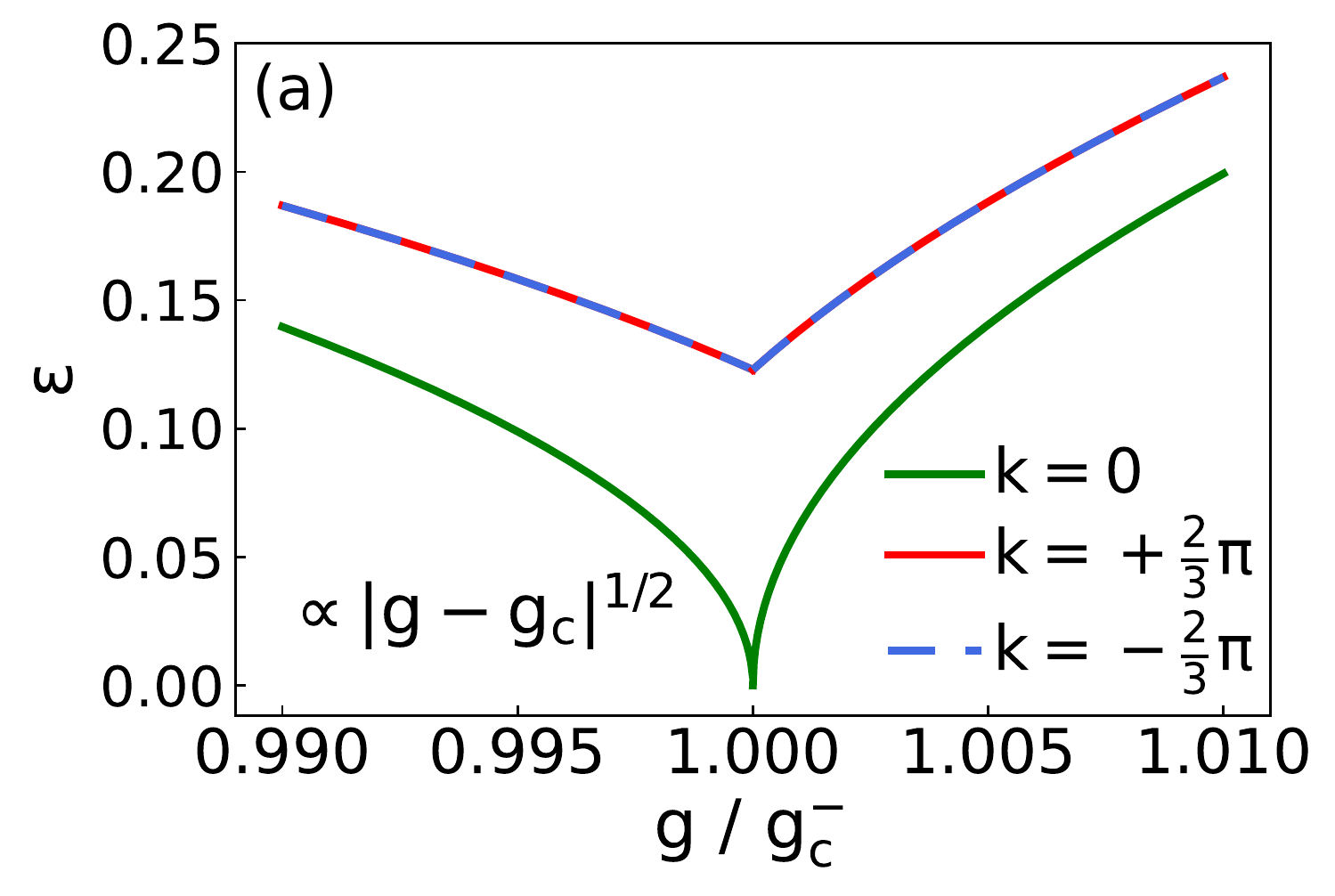} 
\hskip-1ex
\includegraphics[width=0.5\linewidth]{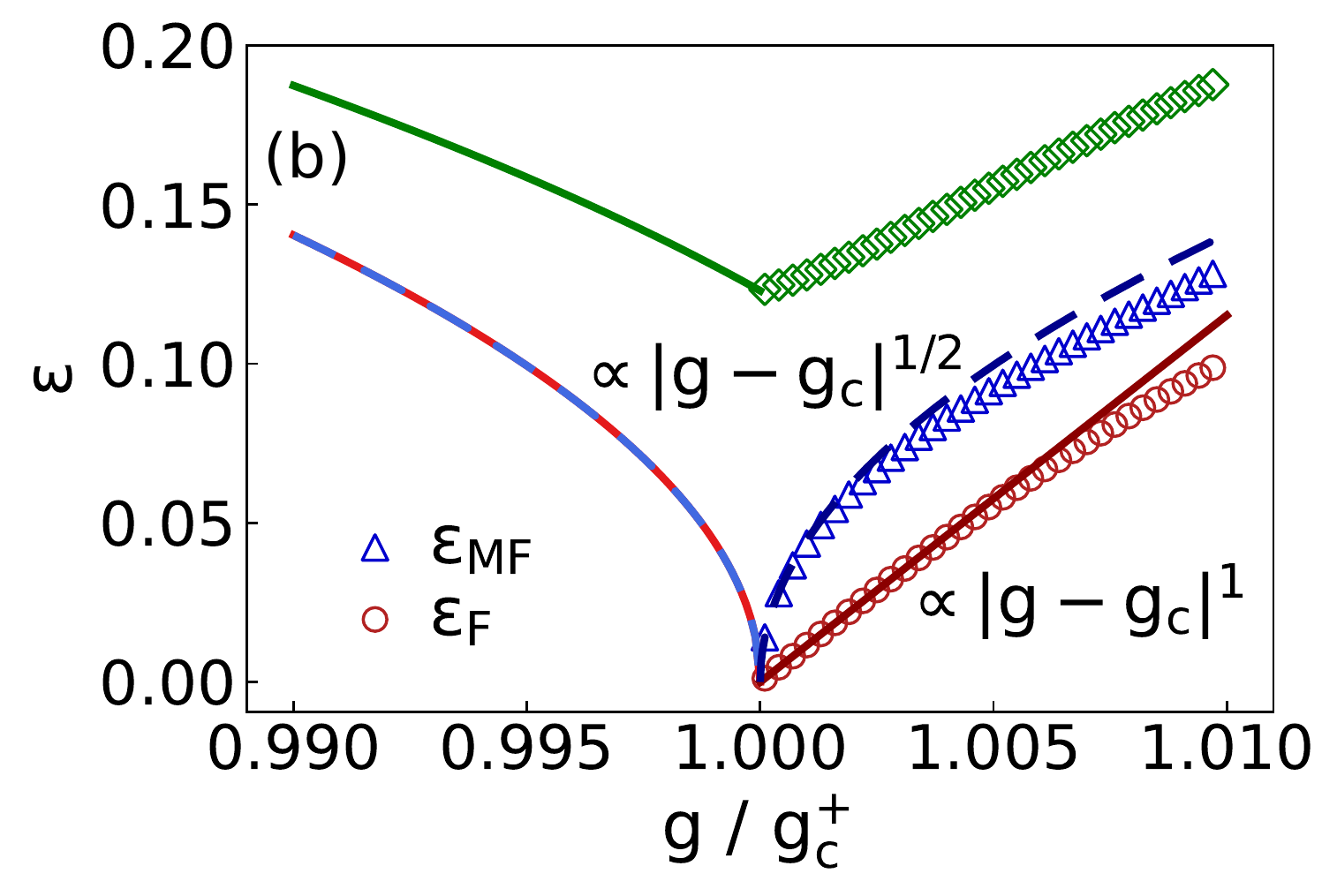} \\
\includegraphics[width=0.5\linewidth]{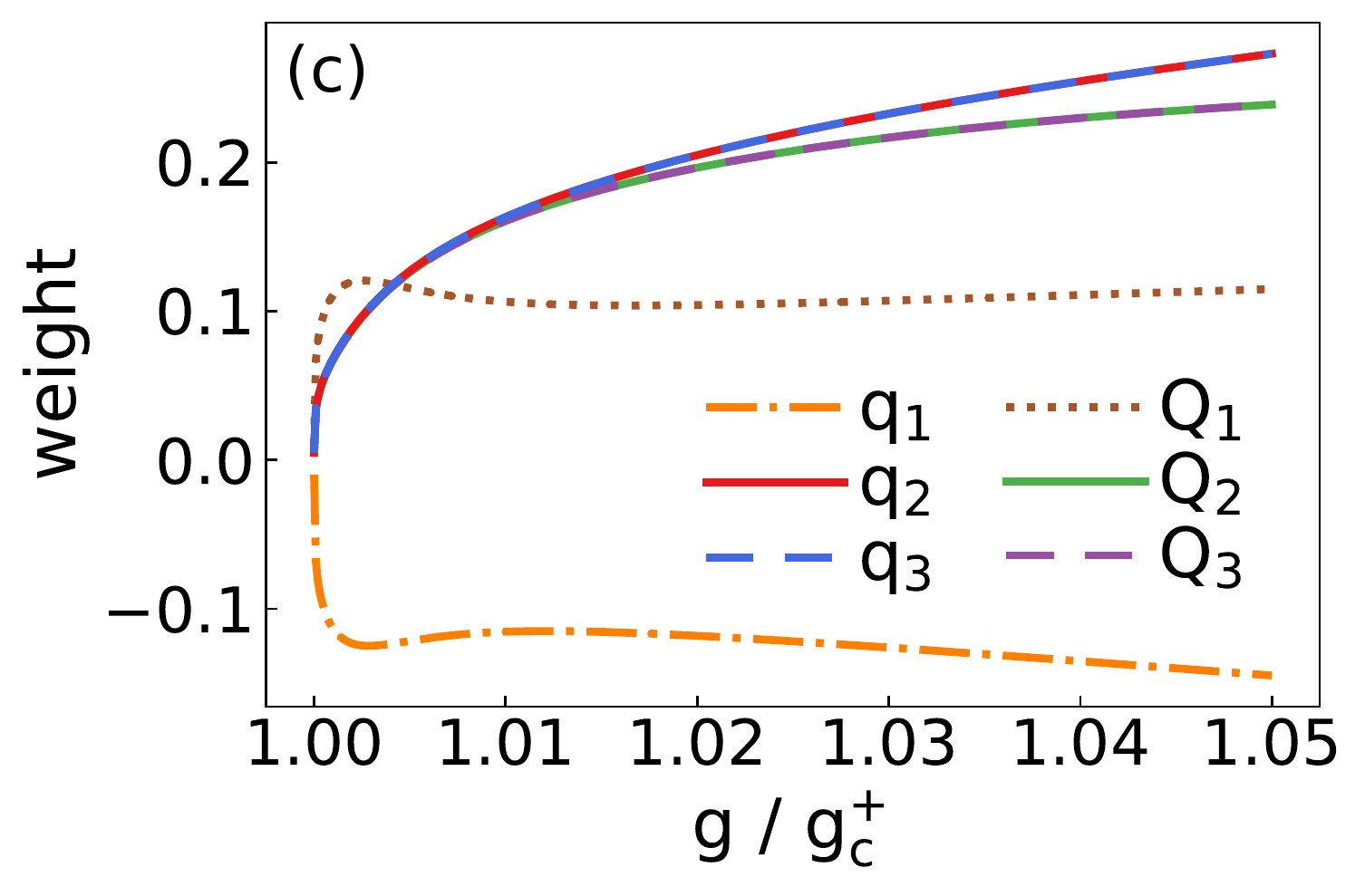} 
\hskip-1ex
\includegraphics[width=0.5\linewidth]{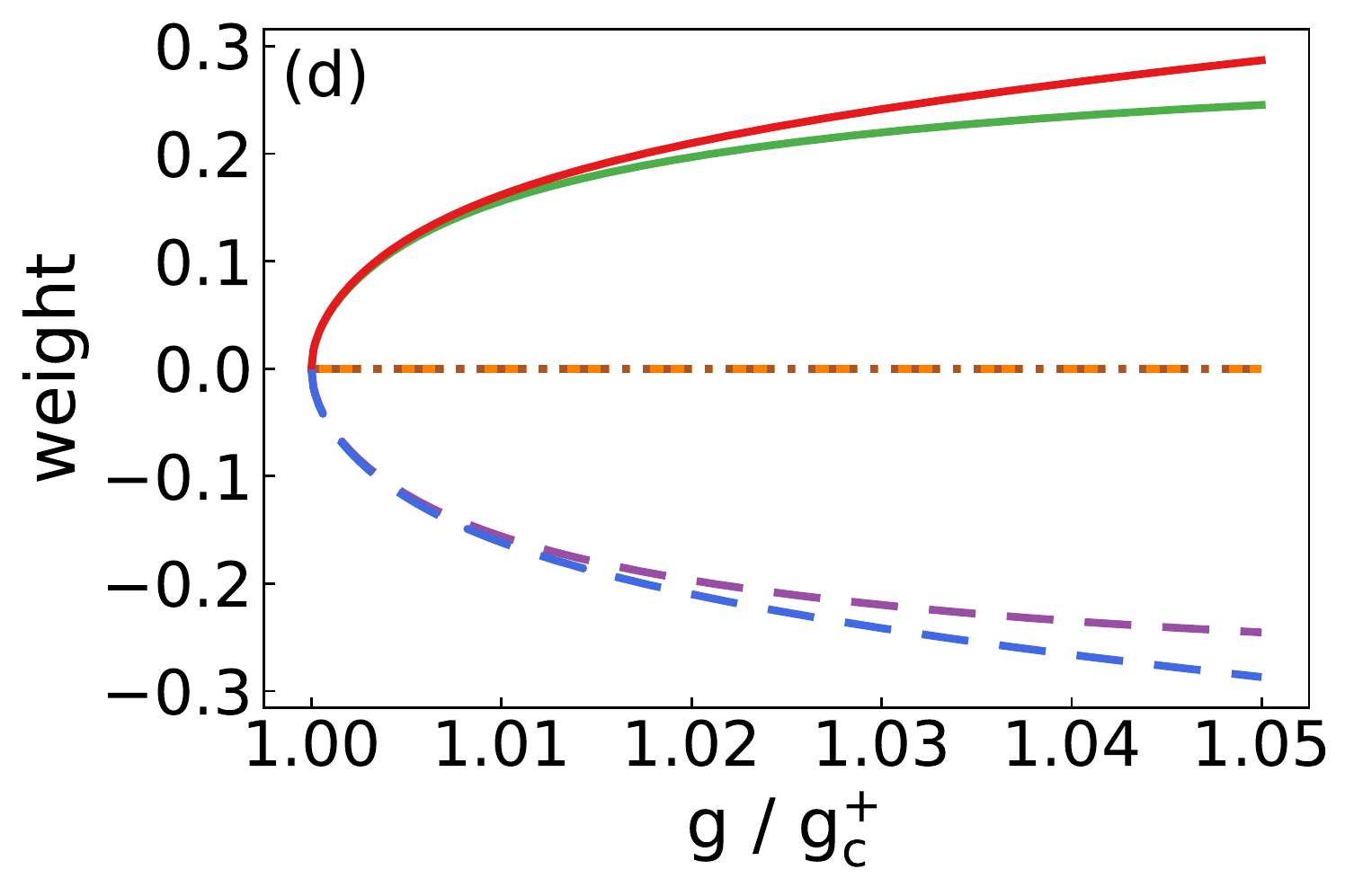}
\caption{Top panel: Excitation energies as a function of the coupling strength $g/g_c^{\pm}$ for (a) $J<0$ and (b) $J>0$. In (b), for $g/g_c^+>1$, shapes and lines are numerical and approximate solutions, respectively, for the two critical modes with distinct power-laws with exponents $\gamma_\textMF=1/2$ and $\gamma_\textF=1$. Bottom panel: The weight of the local fields for (c) the mean-field mode and (d) the frustrated mode for $J>0$ case. The frustrated mode is decoupled completely from the site $1$ as shown in (d). }
\label{fig:2}
\end{figure}

\begin{figure}[t]
\centering
\includegraphics[height=0.35\linewidth]{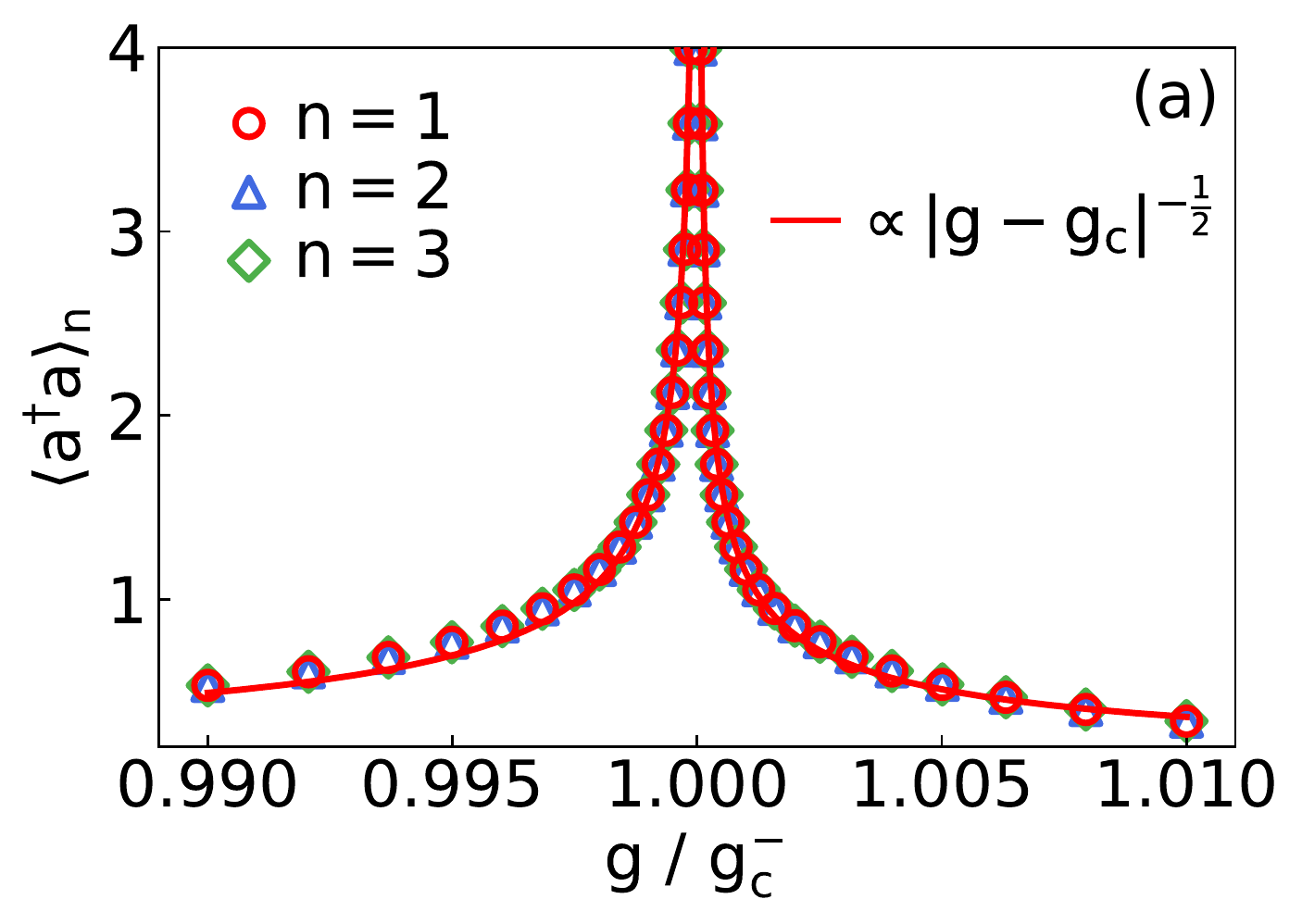}
\hskip-1ex
\includegraphics[height=0.35\linewidth]{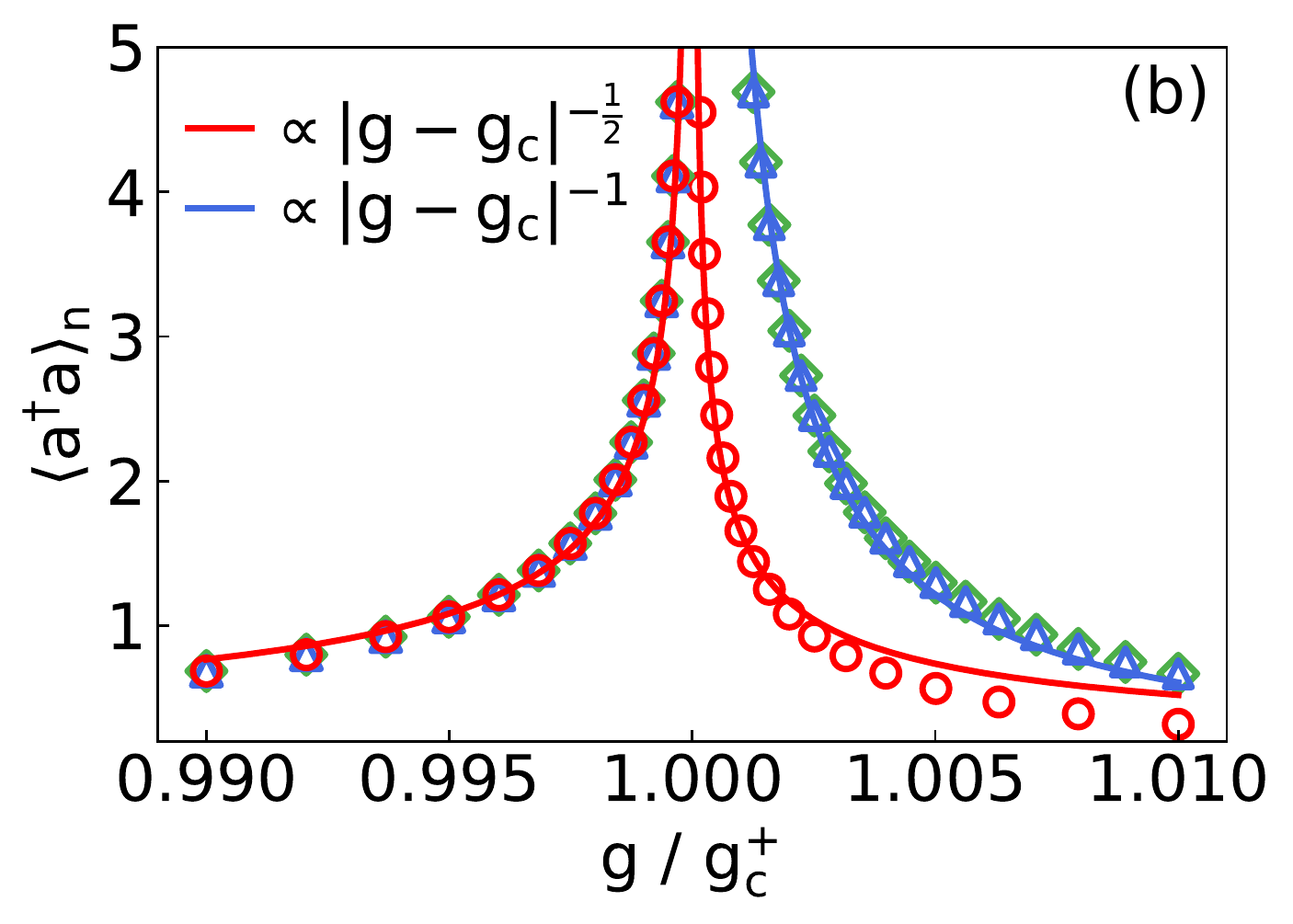} 
\caption{The oscillator population of the ground state as a function of the coupling strength $g/g_c^{\pm}$ for (a) $J<0$ and (b) $J>0$.  Shapes are numerical solutions, and lines indicate corresponding power-laws. (b) For $g>g_c^+$, the photon number of the unpaired site $1$ diverges with a power-law with an exponent $\gamma_\textMF=1/2$, while the remaining sites diverges with an exponent $\gamma_\textF=1$.}
\label{fig:3}
\end{figure}

{\it Excitations and fluctuations with no frustration.---}
Having derived the ground state solution $\{\alpha_n,\theta_n,\phi_m\}$, which gives $H_l=0$, we investigate the excitation spectra and fluctuations using $H_q$ in Eq.~(\ref{eq:transformedH}),
\begin{equation}\label{eq:effhamiltonian}
	\begin{aligned}
		H_q=&\sum_{n=1}^N\Big[\frac{\omega_{0}}{2} \left(q_n^2+p_n^2\right)+\frac{\Omega}{2\cos{\theta_n}} \left(Q_n^2+P_n^2\right)\\
		&+2\lambda\cos{\theta_n}\cos\phi_n Q_n q_n+J(q_{n}q_{n+1}+p_{n}p_{n+1})\Big],
	\end{aligned}
\end{equation}
where $a_n=(q_n+ip_n)/\sqrt{2}$ and $b_n=(Q_n+iP_n)/\sqrt{2}$.
In the normal phase ($g<g_c^{\pm}$) and the NFSP ($g>g_c^{-}$), we apply the discrete Fourier transform $q_{n}=\sum_{k} e^{-i k n} q_{k} / \sqrt{3}$, $Q_{n}=\sum_{k} e^{-i k n} Q_{k} / \sqrt{3}$ with $k=0,~\pm 2 \pi / 3$ to Eq. (\ref{eq:effhamiltonian}) and diagonalize it. For each $k$-momentum mode, there are two branches of excitations (See Supplemental Material~\cite{sup} for analytic expressions), only the lower branch is shown in Fig.~\ref{fig:2} (a) and (b). While $k=0$ mode becomes critical for $J<0$, the degenerate $k=\pm 2 \pi / 3$ modes become critical for $J>0$.  We also calculate the photon number $\braket{a_n^\dagger a_n}$, a fluctuation of the order parameter, and present the result in Fig.~\ref{fig:3}.  In both normal phase and NFSP, we find that the closing energy gap and the diverging photon number, as well as the diverging cavity squeezing $ \braket{q_n^2}-\braket{q_n}^2$, are governed by the mean-field exponent $1/2$~\cite{sup}.

{\it Two critical scalings in the FSP.---} In the FSP, the elementary excitations do not carry a definite momentum because the ground-state solutions given in Eq.~(\ref{eq:minsp2a}) break the translational symmetry. By solving $H_q$~\cite{sup}, we obtain the excitation spectra and the local photon number. The numerical solutions presented in Fig.~\ref{fig:2}(b) and Fig.~\ref{fig:3}(b) exhibit unconventional critical behaviors. First, there are two non-degenerate excitation modes closing their gaps at $g=g_c^+$ with two different power-laws. Second, the local photon numbers diverge on all sites, but their critical exponents are different between the unpaired site ($n=1$) and the ferromagnetic pair ($n=2,3$). The emergence of two co-existing critical diverging time scales (inverse of the excitation energies) and the locally varying critical exponents for fluctuation are striking consequences of the frustrated superradiance and we provide a deeper understanding of these observations below.

By using $\alpha_2=\alpha_3$ for $g>g_c^+$, we derive the lowest excitation energy in the FSP,
\begin{equation}
\label{eq:freqfsp}
	\varepsilon_{\mathrm{F}} = \omega_0\left(A_+-\sqrt{A_-^2-\frac{(\bar\Omega g_c^+g)^2}{\sqrt{1+4g^2\bar\alpha_2^2}}}\right)^{1/2}
\end{equation}
with $2A_\pm=(g_c^+)^4\pm\bar\Omega^2(1+4g^2\bar\alpha_2^2)$, which becomes zero at $g=g_c^+$. Note that it only depends on $\alpha_2=\alpha_3$ (the ferromagnetic pair), but not on $\alpha_1$ (the unpaired site). We denote the second lowest excitation energy which also becomes zero at $g=g_c^+$ as $\varepsilon_\textrm{MF}$. To derive an analytic expression for critical exponents, we use the approximate solution for $\bar\alpha_{1,2,3}$ in Eq.~(\ref{eq:minsp2a}) to expand both $\varepsilon_\textrm{F}$ and $\varepsilon_\textrm{MF}$ around $g\sim g_c^+$ and find
\begin{equation}
	\varepsilon_{\mathrm{F}} \propto|g-g_c^+|^{\gamma_\textF},\quad \varepsilon_{\mathrm{MF}} \propto|g-g_c^+|^{\gamma_\textMF},
\end{equation}
where
\begin{equation}
\gamma_\textF=1,\quad \gamma_\textMF=1/2.
\end{equation}
Here, we denote the combination of $z\nu$~\cite{Emary:2003da,Hwang:2015eq} as $\gamma$, which also serves as the photon number exponent as $\braket{a^\dagger a}\propto 1/\omega$ for a harmonic oscillator~\cite{Torre.2013tfj}. They agree well with the numerical solution [Fig.~\ref{fig:2} (b)]. We refer to one of the critical modes with $\gamma_\textMF$ as a mean-field mode and the other with $\gamma_\textF$ as a frustrated mode. The first term of $\bar\alpha_{2,3}$ in Eq.~(\ref{eq:minsp2a}) makes the coefficient of $|g-g_c^+|^{1/2}$ term in $\varepsilon_\textF$ vanish, whereas the second term leaves the coefficient of $|g-g_c^+|^{1}$ non-zero, leading to $\gamma_\textF=1$~\cite{sup}.

The elementary excitation corresponding to $\varepsilon_{\mathrm{F (MF)}}$ is a superposition of local fields and  can be described by a position quadrature $q_\textrm{F(MF)}=\textbf{v}_\textrm{F(MF)}\cdot\textbf{r}^\top$ with $\textbf{r}=(q_1,Q_1,q_2,Q_2,q_3,Q_3)^\top$ and $\textbf{v}_\textrm{F(MF)}=(v_1,w_1,v_2,w_2,v_3,w_3)^\top$, and a similarly defined momentum quadrature. While the weight vector $\textbf{v}_\textrm{MF}$ has a non-zero support on all sites [Fig.~\ref{fig:2}(c)], $\textbf{v}_\textrm{F}$ has $v_1=w_1=0$, $v_2=-v_3$, and $w_2=-w_3$ [Fig.~\ref{fig:2}(d)]. The latter suggests that the frustrated mode describes the fluctuation of the difference between order parameters of the ferromagnetic pair. This point can be further elucidated by investigating the structure of the mean-field energy landscape in Eq.~(\ref{eq:sqrtE}). For $g>g_c^+$, near frustrated ground-state solutions, we find that two eigenvalues of the Hessian matrix of the energy become zero, $\lambda_\textrm{MF}\propto|g-g_c|^{1}$ and $\lambda_{\textrm{F}}\propto|g-g_c|^{2}$, which in turn give correct critical exponents as $\varepsilon_{\textrm{MF(F)}}\propto\sqrt{\lambda_\textrm{MF(F)}}$ \cite{sup}. The corresponding eigenvector $\textbf{y}_\textrm{F}\propto(0,1,-1)$ of $\lambda_{\textrm{F}}$ is indeed a difference mode between the ferromagnetic pair.

The two critical scalings also appear in the local photon number $\braket{a^\dagger_n a_n}$. As shown in Fig.~\ref{fig:3}(b), we find that the photon numbers of the ferromagnetic pair diverge with $\gamma_\textrm{F}$, while the unpaired site diverges with $\gamma_\textrm{MF}$, namely,
\begin{equation}
\label{photonnumber}
	\langle a^\dagger_1 a_1\rangle\propto|g-g_c|^{-\gamma_\textMF},\quad \langle a^\dagger_2 a_2\rangle=\langle a^\dagger_3 a_3\rangle\propto|g-g_c|^{-\gamma_\textF}.
\end{equation}
The photon number exponent varies locally because the divergence of the local photon number has two contributions, i.e., the diverging excitation numbers of the mean-field mode and the frustrated mode. While the mean-field mode contributes to all three sites with the exponent $\gamma_\textMF$, the frustrated mode contributes only to the ferromagnetic pair with the exponent $\gamma_\textF$, which is the dominant contribution. As the local photon number is readily accessible in cavity QED~\cite{Baumann:2010js} or trapped-ion~\cite{Cai.2021} experiments, Eq.~(\ref{photonnumber}) can serve as a probe for the frustrated SPT and its unconventional scaling behaviors.

\begin{figure}[t]
\centering
\includegraphics[width=0.5\linewidth]{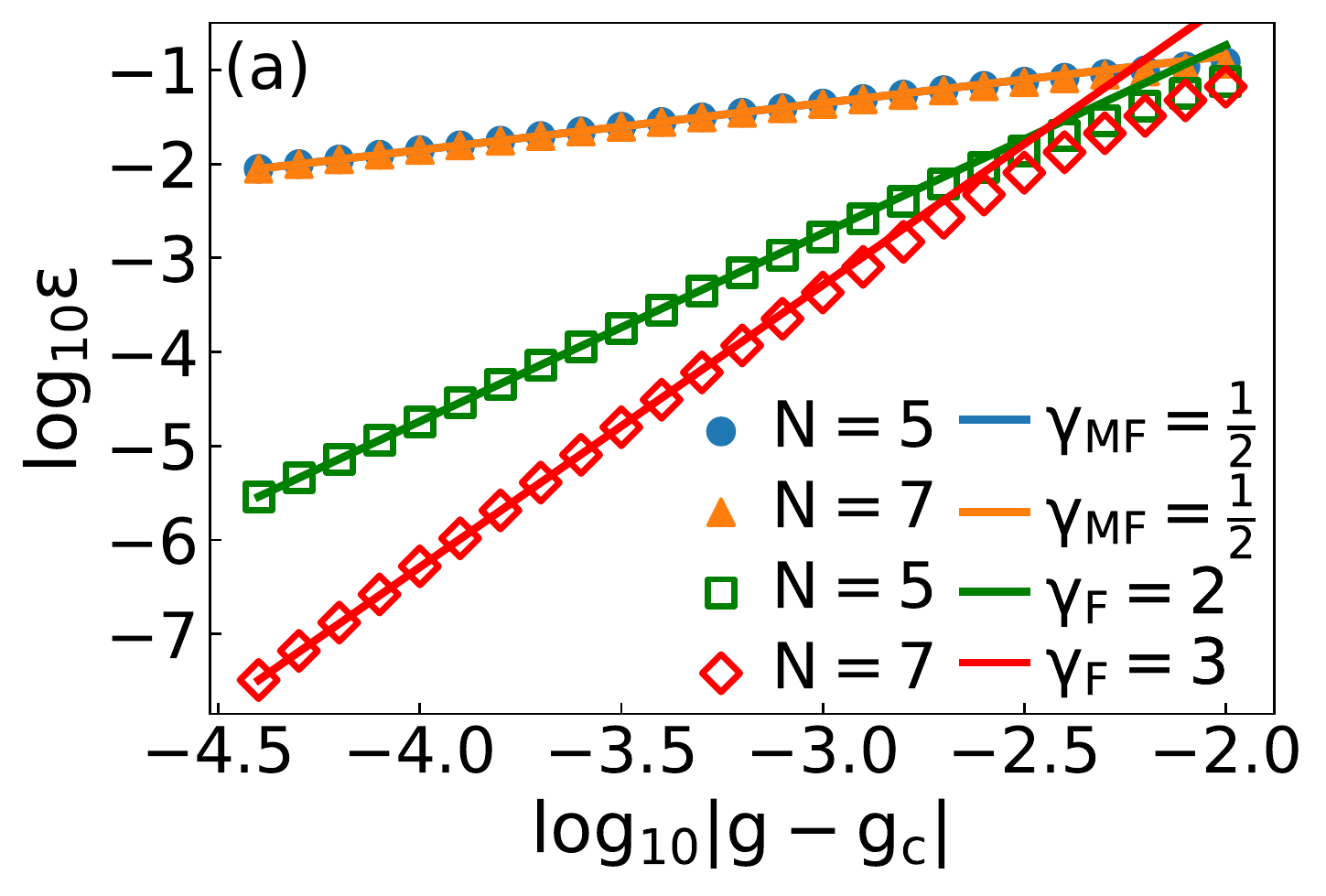}
\includegraphics[width=0.475\linewidth]{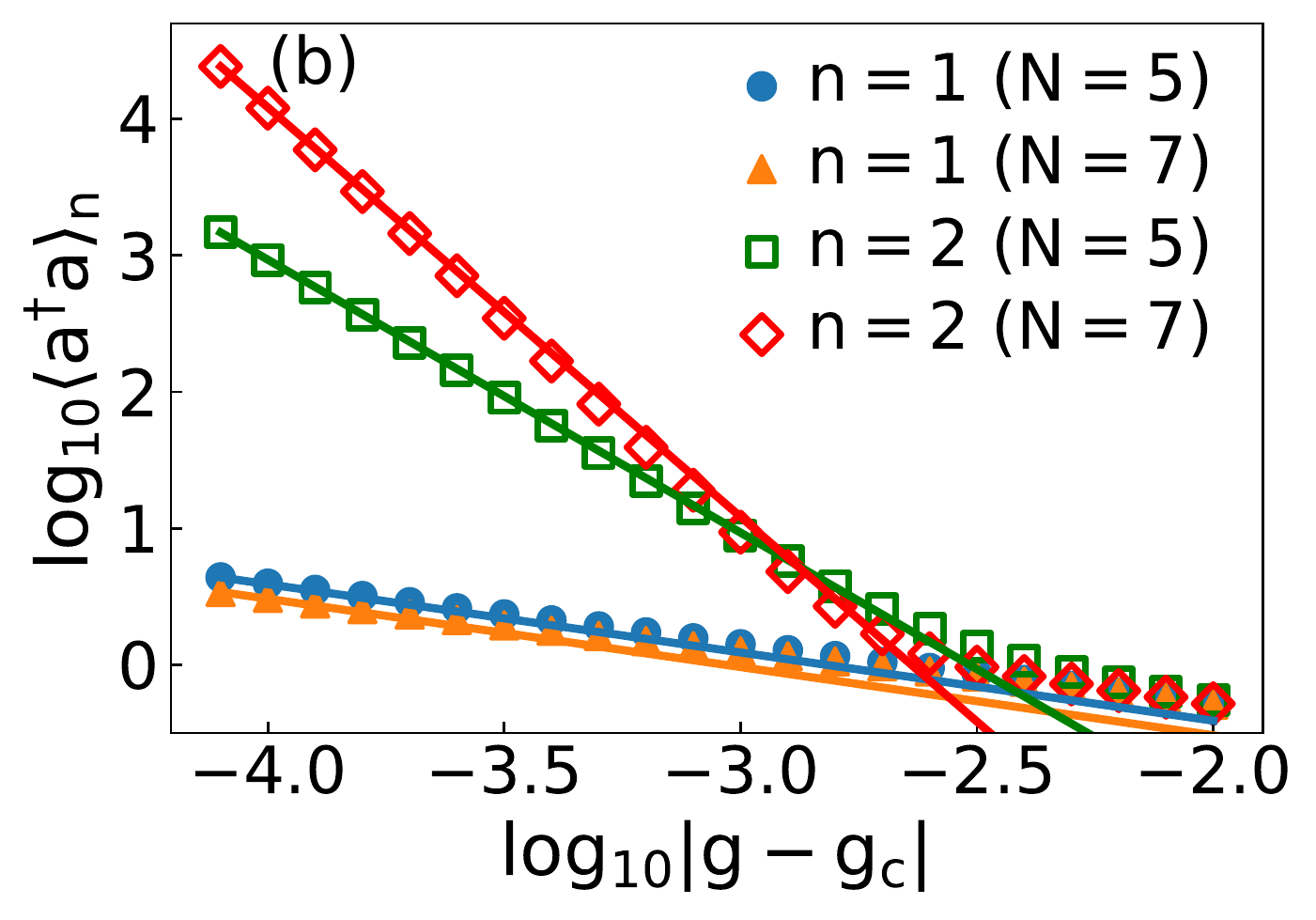} 
\caption{Frustrated superradiant phase of the Dicke lattice with lattice size $N=5$ and $7$. (a) Two critical excitation energies and (b) local photon numbers for the $n=1$ and $2$ cavities. Shapes are numerical solutions and lines indicate corresponding power-laws. The co-existence of two critical scalings is found for any $N$. The mean-field exponent $\gamma_\textMF$ always exists for any $N$, while the frustration critical exponent $\gamma_\textF$ becomes a function of $N$, i.e., $\gamma_\textF(N)=(N-1)/2$.} 
\label{fig:4}
\end{figure}

{\it Frustrated SPTs in lattice systems.---}
The mechanism for the frustrated superradiance we discussed using the triangle motif can be generalized to a higher dimensional system. Here, we consider the simplest example of the $1$D Dicke lattice in Eq.~(\ref{eq:hamiltonian}) with odd $N$ sites to identify generic features of the frustrated SPT and leave the investigation of the $2$D lattice models for future works. We first find that the critical point for $J<0$ is independent of $N$, $g_c^-(N)=\sqrt{1-2\bar J}$; meanwhile, for $J>0$, the critical point depends on $N$,
\begin{equation}
	g_c^{+}(N)=\sqrt{1+2\bar J \cos\left(\frac{N-1}{N}\pi\right)}.
\end{equation}
For $g>g_c^+(N)$,  we find the frustrated ground-state solutions with $2N$-fold degeneracy. As in the Dicke trimer case, there is a single unpaired site with mean-value $\alpha_1$ and $(N-1)/2$ ferromagnetic pairs with the same mean-value $\alpha_{1+j}=\alpha_{N+1-j}$ where $j=1,2,...,(N-1)/2$~\cite{sup}. All neighboring sites are anti-aligned except for $\alpha_{(N+1)/2}=\alpha_{(N+3)/2}$. We numerically calculate the excitation spectra and the local photon numbers and find that the coexistence of two critical scalings, i.e., the mean-field mode and the single frustrated mode, appears for any odd $N$ [See Fig.~\ref{fig:4}]. Moreover, numerical results show that  the frustration critical exponent becomes lattice-size dependent, i.e.,
\begin{equation}
\label{exponent}
\gamma_\textF(N) = (N-1)/2.
\end{equation}
We generalize the insight gained from the Dicke trimer to any odd $N$ sites, namely, the frustrated mode is decoupled from the single unpaired site and describes the fluctuation for the difference in order parameters of ferromagnetic pairs~\cite{sup}. The decoupling of the unpaired site from the frustrated mode explains why the local photon number on the unpaired site diverges with $\gamma_\textMF$ and on all the other sites with $\gamma_\textF(N)$, namely, $\langle a^\dagger a\rangle_1 \propto|g-g_c^+(N)|^{-\gamma_\textMF}$ and $\langle a^\dagger a\rangle_{n\neq 1}  \propto|g-g_c^+(N)|^{-\gamma_\textF(N)}$. The lattice-size dependent critical exponent $\gamma_\textF(N)$ suggests a distinct universality class of the frustrated SPT for each odd $N$.

{\it Physical realizations.---} Our model could be explored in various quantum systems consisting of ubiquitous coupled spins and bosons. The Dicke model could be realized with ion-traps~\cite{Safavi-Naini.2018,Cohn.2018}, superconducting circuits~\cite{Zou:2014cy,Nataf:2010dy,Viehmann:2011dy}, and cavity QED~\cite{Baumann:2010js,Kroeze.2018,Klinder:2015df}, wherein the photon or phonon hopping energy and phase could also be engineered to form a desired lattice~\cite{Kim.2010,Islam:2013ed,Houck:2012iq,Koch:2010eu,Mei.2021,Bermudez.2012si}. Moreover, the proposed mechanism for the frustrated SPT is also applicable to the quantum Rabi model in the limit of $\Omega/\omega_0\rightarrow\infty$~\cite{Hwang:2015eq,Bakemeier:2013hr,Ashhab:2013ke}; this leads to the possibility of finite-component system frustrated SPTs, which could be realized with a small number of qubits and oscillators~\cite{Puebla:2017gq, Cai.2021,Chen.2021}

{\it Conclusion.---}
We have demonstrated that the geometric frustration of the ground-state superradiance of local fields leads to a frustrated SPT and that the emergent FSP exhibits unconventional critical scalings. The frustrated SPT could occur in various lattice geometries and dimensions that go beyond the triangle and 1D lattice considered here and could be realized with a variety of quantum systems. Therefore, our work opens the door for exploring the physics of frustrated superradiant phases both theoretically and experimentally in a wide range of lattice models of coupled bosons and spins.

\begin{acknowledgments}
\emph{Acknowledgments.---}
	The authors thank Lin Jiu for the helpful discussion. This work was supported by NSFC under Grant No. 12050410258 and by Startup Fund and  Summer Research Scholar Program from Duke Kunshan University.	
\end{acknowledgments}

\onecolumngrid

\pagebreak
\widetext
\begin{center}
\textbf{ \large Frustrated Superradiant Phase Transition \\
{\normalsize - Supplemental Material -}}
\end{center}

\setcounter{equation}{0}
\setcounter{figure}{0}
\setcounter{table}{0}

%\makeatletter
\renewcommand{\theequation}{S\arabic{equation}}
\renewcommand{\thefigure}{S\arabic{figure}}
\renewcommand{\bibnumfmt}[1]{[S#1]}
\renewcommand{\citenumfont}[1]{S#1}

\section{I. \quad Effective Hamiltonian in the Thermodynamic Limit}
In this section, we derive the effective Hamiltonian of the Dicke lattice model in the thermodynamic limit $N_a \rightarrow \infty$, given in Eq. (2) in the main text. We will first apply a unitary transformation that separates the mean-field contribution, $\alpha_n=\braket{a_n}$ and $\left(\braket{J_{n}^{x}},\braket{J_{n}^{y}}, \braket{J_{n}^{z}}\right)\rightarrow\frac{N_a}{2}(\sin \theta_n \cos \phi_n, \sin \theta_n \sin \phi_n, \cos \theta_n)$ and the fluctuation around the mean-value, namely,
\begin{equation}
	U(\alpha_n,\theta_n,\phi_n)=\prod_{n=1}^{N}\mathrm{e}^{-i \phi_n J_{n}^z}\mathrm{e}^{-i \theta_n J_n^{y}}\mathrm{e}^{\alpha_n a_{n}^{\dagger}-\alpha_n^*a_{n}},
\end{equation}
to Eq.~(1) in the main text. Here, $\theta_n \in[0,\pi]$ and $\phi_n \in[0,2 \pi)$. The transformed Hamiltonian reads 
\begin{equation}\label{eq:disrot}
 \begin{aligned}
   \tilde{H}_N=U^\dagger H_N U=&\sum_{n=1}^N\bigg[\omega_{0} (a_{n}^{\dagger}+\alpha_n^*) (a_{n}+\alpha_n)+
   \Omega \left(-\sin\theta_n J_{n}^{x}+\cos\theta_n J_{n}^{z}\right)+
   \frac{2\lambda}{\sqrt{N_a}} \left(a_{n}+a_{n}^{\dagger}+2\textrm{Re}[\alpha_n]\right)\times\\
   &\left(\cos\theta_n\cos\phi_nJ_{n}^{x}+ \sin\theta_n\cos\phi_n J_{n}^{z}-\sin\phi_n J_n^y\right)+J\Big((a_{n}^{\dagger}+\alpha_n^*)(a_{n+1}+\alpha_{n+1})+h.c.\Big)\bigg].
 \end{aligned}
\end{equation}
We then apply the Holstein-Primakoff transformation to the collective spin operators, $J_{n}^{+}=\sqrt{N_a-b_{n}^{\dagger} b_{n}}b_{n}$, $\quad J_{n}^{-}=b_{n}^{\dagger} \sqrt{N_a-b_{n}^{\dagger} b_{n}}$, and $J_{n}^{z}=\frac{N_a}{2}-b_{n}^{\dagger} b_{n}$where $b_n$ is a bosonic operator satisfying $\left[b_n, b_n^{\dagger}\right]=1$. In the thermodynamic limit $N_a \rightarrow \infty$, we have $J_{n}^{+}=\sqrt{N_a}b_{n}$, $\quad J_{n}^{-}=\sqrt{N_a}b_{n}^{\dagger}$, and $\quad J_{n}^{z}=\frac{N_a}{2}-b_{n}^{\dagger} b_{n}$.

The ground-state energy consists of terms that depend only on the mean values in $\tilde H_N$, which reads
\begin{equation}\label{eq:EGScomplex}
 E_{\mathrm{GS}} =\sum_{n=1}^N\left[\omega_{0} \alpha_n^{*}\alpha_n + \frac{N_a \Omega}{2}\cos{\theta_n}+{2\lambda}{\sqrt{N_a}}\operatorname{Re}(\alpha_n)\sin{\theta_n}\cos{\phi_n}+J\left(\alpha_n^* \alpha_{n+1} + \alpha_n \alpha_{n+1}^*\right)\right].
\end{equation}
The mean values that minimize the ground-state energy become the ground-state solution. We notice that the energy terms that depend on $\operatorname{Im(\alpha_n)}$ is decoupled from other degrees of freedom $E_{\operatorname{Im}}=\sum_{n=1}^N \left[\omega_0\operatorname{Im}(\alpha_n)^2+2J\operatorname{Im}(\alpha_{n})\operatorname{Im}(\alpha_{n+1})\right]$. Therefore, at minimum energy, we require 	$2\omega_0\operatorname{Im}(\alpha_{n})+2J\operatorname{Im}(\alpha_{n-1})+2J\operatorname{Im}(\alpha_{n+1})=0$, which holds only if $\operatorname{Im(\alpha_n)=0}$.

We can further simplify the ground-state energy by finding the relation between the mean-values of the cavity and atoms (given in Eq. (3) of the main text) and then derive the ground-state energy as a function of cavity fields $\alpha_n$ only, as given in Eq. (4) of the main text. To this end, it is convenient to rescale the energy by $\bar{\alpha}_n=\sqrt{\omega_0/N_a\Omega}~\alpha_n$, $g=2\lambda/\sqrt{\omega_0\Omega}$ to have
\begin{equation}\label{eq:rescaleE}
	\bar{E}_{\mathrm{GS}} = \sum_{n=1}^N\Big(\bar{\alpha}_n^2 + \frac{1}{2}\cos{\theta_n}+g\bar{\alpha}_n\sin{\theta_n}\cos{\phi_n}+2\bar{J}\bar{\alpha}_{n} \bar{\alpha}_{n+1}\Big),
\end{equation}
where $\bar{E}_{\mathrm{GS}}=E_{\mathrm{GS}}/N_a\Omega$ and $\bar{J}=J/\omega_0$. First, since $\sin\theta_n\geq0$, at minimum we have $\cos{\phi_n}=-\bar\alpha_n/|\bar\alpha_n|$; this ensures that the third term has a negative sign with a maximum absolute value. Notice that $\cos\phi_n$ is not defined at $\alpha_n=0$, but it is trivial to $\bar E_{\mathrm{GS}}$ at the origin. Second, by solving $\frac{\partial\bar{E}_{\mathrm{GS}}}{\partial{\theta_n}}=0$, we have
\begin{equation}\label{eq:trigno}
	\sin{\theta_n}={2g|\bar{\alpha}_n}|/{\sqrt{1+4g^2\bar{\alpha}_n^2}},\quad\cos{\theta_n}=-{1}/{\sqrt{1+4g^2\bar{\alpha}_n^2}},
\end{equation}
where the sign of $\cos{\theta_n}$ is fixed to ensure the second term of Eq. (\ref{eq:rescaleE}) is negative. Using the above conditions for $\theta_n$ and $\phi_n$, we finally have the mean-field energy
\begin{equation}
	\bar E_\mathrm{GS}(\{\alpha_n\})= \sum_{n=1}^N\left(\bar{\alpha}_n^2 - \frac{1}{2}{\sqrt{1+4g^2\bar{\alpha}_n^2}}+2\bar{J}\bar{\alpha}_{n} \bar{\alpha}_{n+1}\right),
\end{equation}
which is  Eq. (4) in the main text.

Let us now consider the terms that are linear in bosonic operators,
\begin{equation}
\begin{aligned}
	H_l=\sum_{n=1}^N\left[A_n\left(a_n+a_n^{\dagger}\right)+ B_n \left(b_n+b_n^{\dagger}\right)\right],
\end{aligned}
\end{equation}
where $A_n=\omega_0\alpha_n+J\alpha_{n-1}+J\alpha_{n+1}+{\lambda}{\sqrt{N_a}}\sin\theta_n \cos\phi_n$ and $B_n=-{\sqrt{N_a}}\Omega\sin{\theta_n}+{2\lambda}\alpha_n\cos{\theta_n}\cos{\phi_n}$. Note that both $A_n$ and $B_n$ vanish for the minimum of  $E_\textrm{GS}$, so does $H_l$. Finally, we find that the quadratic term $H_q$ is given by 
\begin{equation}\label{eq:Hq}
  {H}_q = \sum_{n=1}^N\left[\omega_{0} a_{n}^{\dagger} a_{n}-\left(\Omega \cos{\theta_n}+\frac{4\lambda}{\sqrt{N_a}}\alpha_n \sin{\theta_n}\right)b_n^{\dagger}b_n+\lambda\cos{\theta_n}\cos\phi_n \left(a_{n}+a_{n}^{\dagger}\right)\left(b_{n}+b_{n}^{\dagger}\right)+J\left(a_{n}^{\dagger}a_{n+1}+h.c.\right)\right].
\end{equation}
Therefore, the effective Hamiltonian of the Dicke lattice model becomes Eq. (2) in the main text with $H_l=0$, which is exact in the thermodynamic limit where all terms that are inversely proportional to $N_a$ vanish.

\section{II. \quad Mean-field Solution for the Ground State of the Dicke Trimer }

In this section, we find the mean-values that minimize the ground state energy given in Eq. (4) in the main text for the Dicke trimer ($N=3$) with a positive hopping ($J>0$) or a negative hopping ($J<0$) energy. 

\subsection{A. \quad Instability at the origin }

Due to the parity symmetry of the Dicke trimer model, the symmetry preserving ground-state has  zero mean-values, $\alpha_n=0$, which is the origin of parameter space of $(\alpha_1,\alpha_2,\alpha_3)\in\mathbf{R}^3$. The Hessian of $\bar E_{\mathrm{GS}}$ given in Eq. (4) in the main text for $N=3$, is given by 
\begin{equation}\label{eq:hessian}
\textrm{Hess}_3[(\alpha_1,\alpha_2,\alpha_3)]=\left(\begin{array}{ccc}\frac{\partial^{2} \bar{E}}{\partial \bar{\alpha}_1^{2}} & \frac{\partial^{2} \bar{E}}{\partial \bar{\alpha}_1 \partial \bar{\alpha}_2} & \frac{\partial^{2} \bar{E}}{\partial \bar{\alpha}_1 \partial \bar{\alpha}_3} \\ \frac{\partial^{2} \bar{E}}{\partial \bar{\alpha}_2 \partial \bar{\alpha}_1} & \frac{\partial^{2} \bar{E}}{\partial \bar{\alpha}_2^{2}} & \frac{\partial^{2} \bar{E}}{\partial \bar{\alpha}_2 \partial \bar{\alpha}_3} \\ \frac{\partial^{2} \bar{E}}{\partial \bar{\alpha}_3 \partial \bar{\alpha}_1} & \frac{\partial^{2} \bar{E}}{\partial \bar{\alpha}_3 \partial \bar{\alpha}_2} & \frac{\partial^{2} \bar{E}}{\partial \bar{\alpha}_3^{2}}\end{array}\right)=
\left(
\begin{array}{ccc}
 2-\frac{2 g^2}{\left(1+4g^2\bar{\alpha}_1^2\right)^{3/2}} & 2 \bar{J} & 2 \bar{J} \\
 2 \bar{J} & 2-\frac{2 g^2}{\left(1+4g^2\bar{\alpha}_2^2\right)^{3/2}} & 2 \bar{J} \\
 2 \bar{J} & 2 \bar{J} & 2-\frac{2 g^2}{\left(1+4g^2\bar{\alpha}_3^2\right)^{3/2}} \\
\end{array}
\right).
\end{equation}
At the origin, we have
\begin{equation}
\textrm{Hess}_3[(0,0,0)]=\left(
\begin{array}{ccc}
 2-2 g^2 & 2 \bar{J} & 2 \bar{J} \\
 2 \bar{J} & 2-2 g^2 & 2 \bar{J} \\
 2 \bar{J} & 2 \bar{J} & 2-2 g^2 \\
\end{array}
\right),
\end{equation}
whose eigenvalues are $2\left(1-\bar{J}-g^{2}\right)$, $2\left(1-\bar{J}-g^{2}\right)$, and $2\left(1+2 \bar{J}-g^{2}\right)$. Therefore, the origin becomes unstable if $g>\operatorname{min}\left\{\sqrt{1-\bar{J}},\sqrt{1+2\bar{J}}\right\}$, which indicates the emergence of the superradiant solutions for $g>g_c^+\equiv\sqrt{1-\bar{J}}$ for $J>0$ and $g>g_c^-\equiv\sqrt{1+2\bar{J}}$ for $J<0$.

Note that $g_c^\pm$ becomes imaginary for $\bar{J}<-1/2$ and $\bar J>1$. This is because the coupled harmonic oscillators become unstable when the hopping rates become comparable or larger than the oscillator frequency. To find the range of stability for the normal phase, we set $\alpha_n=0$ and $\lambda=0$ in Eq. (\ref{eq:Hq}) and then apply a Fourier transform $a_{n}^{\dagger}=\sum_{k} e^{i kn} a_{k}^{\dagger} / \sqrt{3},b_{n}^{\dagger}=\sum_{k} e^{i kn} b_{k}^{\dagger} / \sqrt{3}$ with $k=0,\pm 2\pi/3$ 
\begin{equation}
 H_q=\sum_{k}\left[(\omega_0+2J\cos k) a_{k}^{\dagger}a_{k}+\Omega b_k^{\dagger}b_k\right].
\end{equation}
Therefore, the Dicke lattice model becomes unstable for $\bar{J}<-1/2$ and $\bar J>1$ as one of the normal mode frequencies, $\omega_0+2J\cos k$, become negative. In the main text, we consider only small hopping rates $-1/2<\bar{J}<1$, where the model is stable. One can see that the atomic part remains always stable and therefore the range of stability is identical for the Rabi lattice model.

\subsection{B. \quad Frustrated superradiant solution ($J>0$)}

For $J>0$, the ground state energy in the Eq. (4) of the main text is minimized when only one neighboring pair takes the same sign and all the other pairs have opposite signs. For $N=3$, therefore, we assume without loss of generality that $\bar{\alpha}_1\leq0$ and $\bar{\alpha}_2,\bar{\alpha}_3\geq0$, which includes the origin. The first order partial derivatives of Eq. (4) from the main text must vanish at local extreme points, i.e.,
\begin{equation}\label{eq:sqrtpd}
	2\bar{\alpha}_{n-1}+2\bar{J} \bar{\alpha}_{n}+2\bar{J} \bar{\alpha}_{n+1}-\frac{2 g^{2} \bar{\alpha}_{n}}{\sqrt{1+4g^{2} \bar{\alpha}_{n}^{2}}}=0,
\end{equation}
which we rearrange to get
\begin{equation}
	\bar{J} (\bar{\alpha}_{n-1}+\bar{\alpha}_{n}+\bar{\alpha}_{n+1})=\frac{g^{2} \bar{\alpha}_{n}}{\sqrt{1+4g^{2} \bar{\alpha}_{n}^{2}}}+(\bar{J}-1) \bar{\alpha}_{n}.
\end{equation}
Therefore, $\bar{\alpha}_{1,2,3}$ that satisfy the above equation can be considered as the roots of $f(\bar{\alpha})=k$, where
\begin{equation}\label{eq:faeqk}
 f(\bar{\alpha})=\frac{g^2\bar{\alpha}}{\sqrt{1+4g^2\bar{\alpha}^2}}+(\bar{J}-1)\bar{\alpha},\quad k=\bar{J}(\bar{\alpha}_1+\bar{\alpha}_2+\bar{\alpha}_3).
\end{equation}
In the following, we will use this condition for the extreme points to prove that i) for $g<g_c^+$, the origin is the global minimum and ii) for $g>g_c^+$, the neighboring pair with the same sign in the frustrated solution is in fact identical, i.e., $\bar\alpha_2=\bar\alpha_3$.

\subsubsection{1. \quad A proof for the origin being the global minimum for  $g<g_c^+$}
For $g<g_c^+$, $f(\bar{\alpha})$ is a monotonously decreasing function that passes through the origin as shown in Fig. \ref{fig:fa} (a). From this, we can show that the only possible extreme point is the origin: if we have $\bar{\alpha}_1+\bar{\alpha}_2+\bar{\alpha}_3<0$ and thus $\bar{J}(\bar{\alpha}_1+\bar{\alpha}_2+\bar{\alpha}_3)<0$, then the solution of Eq. (\ref{eq:faeqk}) will satisfy $\bar{\alpha}_1=\bar{\alpha}_2=\bar{\alpha}_3>0$ (indicated by the red dashed line), which contradicts with our assumption. Similarly, $\bar{\alpha}_1+\bar{\alpha}_2+\bar{\alpha}_3>0$ is also not possible. Therefore, we conclude that $y=f(\bar{\alpha})$ and $y=\bar{J}(\bar{\alpha}_1+\bar{\alpha}_2+\bar{\alpha}_3)$ can only intersect at the origin, 
\begin{equation}\label{eq:npmin2}
 \phi_n=\bar{\alpha}_n=0,\quad\theta_n=\pi,
\end{equation}
with ground-state energy $\bar E_{\mathrm{GS}}^{\mathrm{np}}=-{3}/{2}$. As the origin is the only extreme point, together with the Hessian analysis in Sec. II-A, we conclude that it is a global minimum.

\subsubsection{2. \quad A proof that $\bar\alpha_2=\bar\alpha_3$ for the frustrated solution for $g>g_c^+$.}

\begin{figure}[b]
\centering
{{\includegraphics[width=0.35\textwidth]{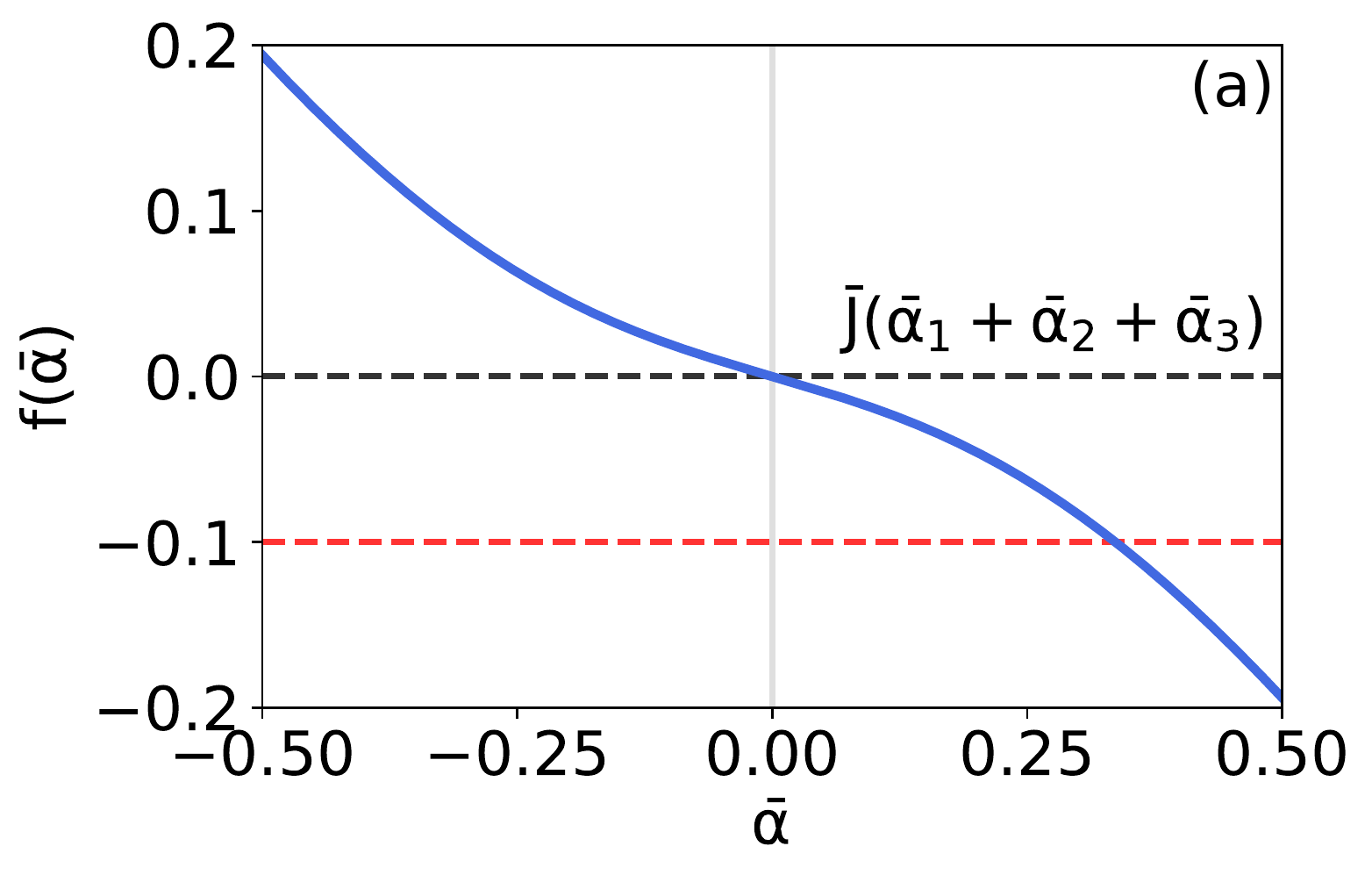} }}
{{\includegraphics[width=0.35\textwidth]{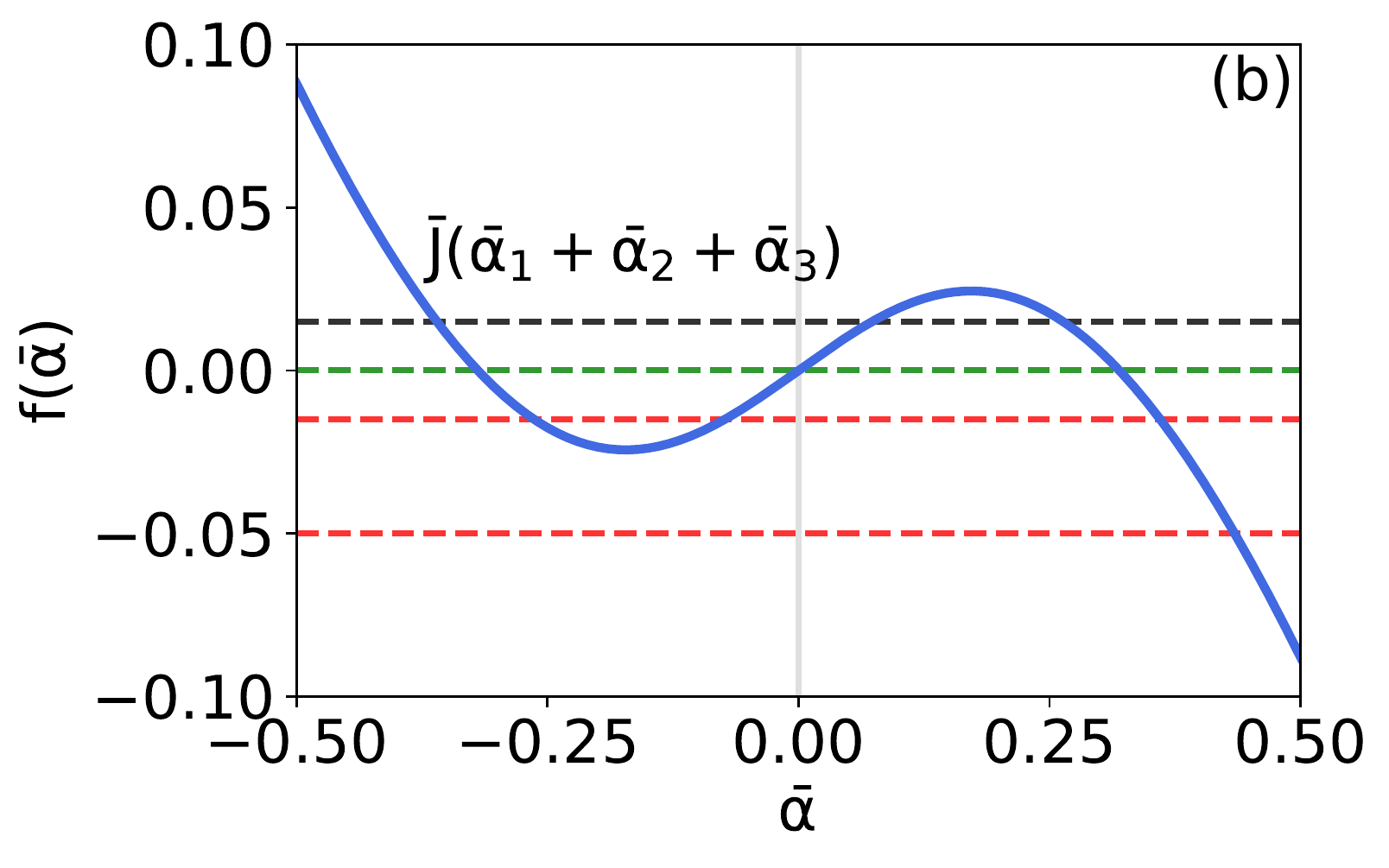} }}
\caption{Monotonicity of function $f(\bar{\alpha})$ for (a) $g<\sqrt{1-\bar{J}}$ and (b) $g>\sqrt{1-\bar{J}}$. The blue line displays the function $f(\bar{\alpha})$, black dashed lines mark the possible solutions for the local minimum, the green line marks the solution for a saddle point, and red dashed lines indicate impossible solutions for an extreme point.}
\label{fig:fa}
\end{figure}

For $g>g_c^+$, $f(\bar{\alpha})$ becomes non-monotonic as shown in Fig. \ref{fig:fa} (b). Our strategy is as follows: (a) We first show that there are no extreme points satisfying $\bar{\alpha}_1+\bar{\alpha}_2+\bar{\alpha}_3<0$. (b) We then prove that there exists an extreme point giving rise to $\bar{\alpha}_1+\bar{\alpha}_2+\bar{\alpha}_3=0$, i.e., ${\alpha}_3=0,{\alpha}_1=-{\alpha}_2$, but we show that this is a saddle point. (c) Finally, we show that $\bar{\alpha}_1<0$ and $\bar\alpha_2=\bar\alpha_3$ is the only allowed solution that leads to $\bar{\alpha}_1+\bar{\alpha}_2+\bar{\alpha}_3>0$.

For the point (a), if one assumes $\bar{\alpha}_1+\bar{\alpha}_2+\bar{\alpha}_3<0$, there are two possibilities as indicated by two red dashed lines in Fig. \ref{fig:fa}(b). The lower red dashed line cannot be satisfied by the same reason for $g<g_c^+$ case. For the upper red dashed line case,  the only allowed combination is $\bar{\alpha}_2=\bar{\alpha}_3>\bar\alpha_{\mathrm{zero}}$ and $\bar{\alpha}_1>-\bar\alpha_{\mathrm{zero}}$, where $\bar\alpha_{\mathrm{zero}}$ is defined by $f(\bar\alpha_{\mathrm{zero}})=0$ and $\bar\alpha_{\mathrm{zero}}>0$. It then leads to $\bar{\alpha}_1+\bar{\alpha}_2+\bar{\alpha}_3>\bar\alpha_{\mathrm{zero}}>0$ which contradicts the assumption $\bar{\alpha}_1+\bar{\alpha}_2+\bar{\alpha}_3<0$. Therefore, we conclude that there is no solution satisfying $\bar{\alpha}_1+\bar{\alpha}_2+\bar{\alpha}_3<0$.

For the point (b), we assume $\bar{\alpha}_1+\bar{\alpha}_2+\bar{\alpha}_3=0$, which is indicated as a green dashed line Fig. \ref{fig:fa}(b). There is an allowed solution satisfying this condition, namely,
\begin{equation}\label{eq:a1a2}
	 \bar{\alpha}_3=0,\quad\bar{\alpha}_{2}=-\bar{\alpha}_{1}=\frac{1}{2 g}{\sqrt{\left(\frac{g}{g_c^+}\right)^{4}-1}},
\end{equation}
which is therefore an extreme point. The Hessian given in Eq. (\ref{eq:hessian}) at this point Eq. (\ref{eq:a1a2}) becomes

\begin{equation}
 \textrm{Hess}_3\left[\left(-\frac{1}{2 g}\sqrt{\left(\frac{g}{g_c^+}\right)^{4}-1},\frac{1}{2 g}\sqrt{\left(\frac{g}{g_c^+}\right)^{4}-1},0\right)\right]=
\left(
\begin{array}{ccc}
 2-\frac{2 (1-\bar {J})^3}{g^4} & 2 \bar {J} & 2 \bar {J} \\
 2 \bar {J} & 2-\frac{2 (1-\bar {J})^3}{g^4} & 2 \bar {J} \\
 2 \bar {J} & 2 \bar {J} & 2-2 g^2 \\
\end{array}
\right),
\end{equation}
whose eigenvalues are given by
\begin{equation}\label{eq:beta}
	\beta_1=2(A_2-\bar J), \quad \beta_{2(3)}=A_1+A_2+\bar J \pm \sqrt{(A_2-A_1)^2+2 \bar J (A_2-A_1)+9 \bar J^2}.
\end{equation}
where $A_1=2-2 g^2,A_2=2-2 (1-\bar {J})^3/g^4$. Note that $\beta_1,\beta_2>0$ and $\beta_3<0$ for $g>g_c^+$ and the point in Eq. (\ref{eq:a1a2}) is therefore a saddle point. 

Finally, the only possibility is then $\bar{\alpha}_1+\bar{\alpha}_2+\bar{\alpha}_3>0$. The case indicated by the top black dashed line in Fig. \ref{fig:fa}(b) is the only solution consistent with our assumption. This leads to $\bar{\alpha}_1<0$ and $\bar{\alpha}_2,\bar{\alpha}_3>0$, which allows two types of solutions, i.e., $\bar\alpha_1<0<\bar\alpha_2=\bar\alpha_3$ and $\bar\alpha_1<0<\bar\alpha_2<\bar\alpha_3$. We provide a proof below that the latter contradicts with $\bar{\alpha}_1+\bar{\alpha}_2+\bar{\alpha}_3>0$, leaving the former as the only possibility.

\begin{proof}
Suppose we have $\bar{\alpha}_1+\bar{\alpha}_2+\bar{\alpha}_3>0$ and $\bar\alpha_1<0<\bar\alpha_2<\bar\alpha_3$. Eq. (\ref{eq:faeqk}) can be rewritten as
\begin{equation}\label{eq:fandk2}
	g^2 \bar\alpha = \left[k+(1- \bar J)\bar\alpha\right]\sqrt{1+4g^2\bar\alpha^2}.
\end{equation}
Since $\sqrt{1+4g^2\bar\alpha^2}>0$, we require $\bar\alpha$ to satisfy 
\begin{equation}\label{eq:3root}
	g^2\bar\alpha/\left[k+(1-\bar J)\bar\alpha\right]>0.
\end{equation}
Squaring both sides of Eq. (\ref{eq:fandk2}) gives
\begin{equation}\label{eq:quatic}
	g^4 \bar\alpha^2 = \left[k+(1- \bar J)\bar\alpha\right]^2(1+4g^2\bar\alpha^2).
\end{equation}
The sum of roots for a quartic equation $c_4 x^{4}+c_3 x^{3}+c_2 x^{2}+c_1 x+c_0=0$ is given by $-c_3/c_4$. In Eq. (\ref{eq:quatic}), we have $c_4 = 4g^2(1- \bar J)^2,~c_3=8g^2k(1- \bar J)$. 
Therefore, one finds $\bar\alpha_1+\bar\alpha_2+\bar\alpha_3+\bar\alpha_4=-{c_3}/{c_4}=-{2k}/(1- \bar J)<0$. From our assumption $\bar\alpha_1+\bar\alpha_2+\bar\alpha_3>0$, we conclude that, by squaring Eq. (\ref{eq:fandk2}), a new root $\bar\alpha_4<0$ is added to Eq. (\ref{eq:quatic}), which is not a solution of Eq. (\ref{eq:fandk2}). Hence, $\alpha_4$ must contradict Eq. (\ref{eq:3root}) and satisfy
\begin{equation}
	k+(1- \bar J)\bar\alpha_4>0 \Rightarrow \bar\alpha_4>-\frac{k}{1- \bar J}.
\end{equation}
Using this bound for $\alpha_4$ together with the sum of all four roots, we find
\begin{equation}
	\bar\alpha_1+\bar\alpha_2+\bar\alpha_3 = -\frac{2k}{1- \bar J} - \bar\alpha_4 < -\frac{k}{1- \bar J}<0,
\end{equation}
which contradicts $\bar{\alpha}_1+\bar{\alpha}_2+\bar{\alpha}_3>0$. Therefore, $0<\bar\alpha_2<\bar\alpha_3$ is not an allowed solution. We briefly remark on the role of $\bar\alpha_2<\bar\alpha_3$ in this proof. It ensures $\{\bar\alpha_1,\bar\alpha_2,\bar\alpha_3\}$ corresponds to three distinct roots of $f(\bar\alpha)=k$, so we can apply the sum of roots of Eq. (\ref{eq:quatic}). On the other hand, if $\bar\alpha_2=\bar\alpha_3$, $\{\bar\alpha_1,\bar\alpha_2,\bar\alpha_3\}$ consists of two of the three roots of $f(\bar\alpha)=k$ [as long as $\bar\alpha_2=\bar\alpha_3$ is not a multiple root at the extreme point of $f(\bar\alpha)$, which can be easily verified by partial derivatives]; in this case, the sum of $\alpha_n$ is no longer the sum of roots, and therefore the above contradiction no longer holds.
\end{proof}

\subsubsection{3. \quad Approximate Solution}
Using $\bar{\alpha}_2=\bar{\alpha}_3$ for the global minimum, we can reduce Eq. (\ref{eq:sqrtpd}) to
\begin{subequations}\label{eq:sqrtpd3}
\begin{align}
\bar{\alpha}_{1}+2\bar{J} \bar{\alpha}_{2}-\frac{g^{2} \bar{\alpha}_{1}}{\sqrt{1+4g^{2} \bar{\alpha}_{1}^{2}}}=0, \label{eq:sqrtpd31}\\
 (1+\bar{J})\bar{\alpha}_{2}+\bar{J} \bar{\alpha}_{1}-\frac{g^{2} \bar{\alpha}_{2}}{\sqrt{1+4g^{2} \bar{\alpha}_{2}^{2}}}=0. \label{eq:sqrtpd32}
\end{align}
\end{subequations}
While the above coupled equations cannot be solved analytically, we find approximate solutions for $\alpha_n$  for $|g-g_c^+|\ll1$, which reads
\begin{subequations}%\label{eq:minsp2a}
	\begin{align}
	\bar{\alpha}_{1}& \simeq -\frac{2 |g-g_c^+|^{1/2}}{\sqrt{3} (g_c^+)^{3/2}} - \frac{|g-g_c^+|^{3/2}}{6 \sqrt{3} (g_c^+)^{5/2}}+O\left(\left(g-g_c^+\right)^{5/2}\right),\\
		\bar{\alpha}_{2}=\bar{\alpha}_{3}& \simeq \frac{|g-g_c^+|^{1/2}}{\sqrt{3} (g_c^+)^{3/2}} +\frac{(8-7\bar{J}) |g-g_c^+|^{3/2}}{12\sqrt{3}\bar{J} (g_c^+)^{5/2}}+O\left(\left(g-g_c^+\right)^{5/2}\right),
	\end{align}
\end{subequations}
which is Eq. (6) of the main text. Using Eq.~(\ref{eq:trigno}), we find the mean-values for atoms up to the same order in $|g-g_c^+|\ll1$ as
\begin{subequations}\label{eq:Ftheta}
\begin{align}
 \cos{\theta_1}& \simeq -1+\frac{8 \left(g-g_c^+\right)}{3 g_c^+}-\frac{44 \left(g-g_c^+\right)^2}{9(g_c^+)^2}+O\left(\left(g-g_c^+\right)^3\right),\\
 \cos{\theta_2}=\cos{\theta_3}&\simeq-1+\frac{2 \left(g-g_c^+\right)}{3 g_c^+}+\frac{(8-\bar{J}) \left(g-g_c^+\right)^2}{9 \bar{J}(g_c^+)^2}+O\left(\left(g-g_c^+\right)^3\right).
\end{align}
\end{subequations}
Moreover, $\phi_1=0$ and $\phi_2=\phi_3=\pi$. The rescaled ground-state energy near the critical point therefore is given by
\begin{equation}\label{eq:gs_sp2_exp}
\bar E_{\mathrm{GS}}^{\mathrm{fsp}}\simeq-\frac{3}{2}-\frac{2 \left(g-g_c^+\right)^2}{(g_c^+)^2}+O\left(\left(g-g_c^+\right)^3\right)
\end{equation}

\subsection{C. \quad Non-frustrated superradiant solution ($J<0$)}

For $J<0$,we find an exact analytic solution of the mean-values. To this end, we use following inequalities,
\begin{subequations}\label{eq:inequalities}
 \begin{align}
  \sum_{i=1}^3\sqrt{1+4g^2\bar{\alpha}_i^2}&
  \leq3\sqrt{1+\frac{4}{3}g^2\left(\sum_{i=1}^3\Bar{\alpha}_i^2\right)},\\
  \Bar{\alpha}_1 \Bar{\alpha}_2 + \Bar{\alpha}_1 \Bar{\alpha}_3 + \Bar{\alpha}_2 \Bar{\alpha}_3 &\leq \Bar{\alpha}_1^2+\Bar{\alpha}_2^2+\Bar{\alpha}_3^2,
 \end{align}
\end{subequations}
where the equalities hold if and only if $\Bar{\alpha}_1^2=\Bar{\alpha}_2^2=\Bar{\alpha}_3^2$ and $\Bar{\alpha}_1=\Bar{\alpha}_2=\Bar{\alpha}_3$, respectively. The first inequality is the Cauchy-Schwarz inequality (AM-QM inequality). Using Eq.~(\ref{eq:inequalities}) and $\bar J<0$, we find the lower-bound for the ground-state energy
\begin{equation}\label{eq:Eineq3}
\begin{aligned}
 \Bar{E}_{\mathrm{GS}} & \geq (\Bar{\alpha}_1^2+\Bar{\alpha}_2^2+\Bar{\alpha}_3^2) - \frac{3}{2} \sqrt{1+\frac{4}{3}g^2(\Bar{\alpha}_1^2+\Bar{\alpha}_2^2+\Bar{\alpha}_3^2)}+2\Bar{J}(\Bar{\alpha}_1 \Bar{\alpha}_2 + \Bar{\alpha}_1 \Bar{\alpha}_3 + \Bar{\alpha}_2 \Bar{\alpha}_3)\\
 & \geq (1+2\bar{J})(\Bar{\alpha}_1^2+\Bar{\alpha}_2^2+\Bar{\alpha}_3^2) - \frac{3}{2} \sqrt{1+\frac{4}{3}g^2(\Bar{\alpha}_1^2+\Bar{\alpha}_2^2+\Bar{\alpha}_3^2)},
\end{aligned}
\end{equation}
where the equality holds if and only if $\Bar{\alpha}_1=\Bar{\alpha}_2=\Bar{\alpha}_3$. Therefore, the minimum ground-state energy can be obtained by minimizing
\begin{equation}
    \Bar{E}_{\mathrm{GS}}= 3(1+2\bar{J})\Bar{\alpha}^2 - \frac{3}{2} \sqrt{1+4g^2\Bar{\alpha}^2}=3\left((g_c^-\bar\alpha)^2 - \frac{1}{2}\sqrt{1+4(g/g_c^-)^{2}(g_c^-\bar\alpha)^2}\right).
\end{equation}
%. and the results are identical with the single dicke with renormalized parameters.
where $\bar\alpha\equiv	\bar{\alpha}_1=\bar{\alpha}_2=\bar{\alpha}_3$. Note that in the second equality, the ground-state energy can be expressed in the same functional form with the single Dicke model. This shows that the nature of superradiant phase transition for the Dicke trimer with $J<0$ is identical to that of the single Dicke model, with the only change being the shift of the critical point. It therefore follows that for $g<g_c^-$, the global minimum locates at 
\begin{equation}\label{eq:npmin1}
\bar{\alpha}_1=\bar{\alpha}_2=\bar{\alpha}_3=0,
\end{equation}
with the ground-state energy $\bar E_{\mathrm{GS}}^{\mathrm{np}}=-3/2$. For $g>g_c^-$, the global minimum locates at
\begin{equation}
\bar{\alpha}_1=\bar{\alpha}_2=\bar{\alpha}_3=\frac{1}{2g}\sqrt{\left(\frac{g}{g_c^-}\right)^4-1}.
\end{equation}
Note that $\bar\alpha_n$ can also be negative, indicating a two-fold degenerate ground state energy,
\begin{equation}\label{eq:gs_sp1}
\bar E_{\mathrm{GS}}^{\mathrm{nfsp}} = -\frac{3}{4}\left[\frac{(g_c^-)^2}{ g^2}+\frac{g^2}{(g_c^-)^2}\right].
\end{equation}

\begin{figure}[t]
\centering
\includegraphics[height=0.20\linewidth]{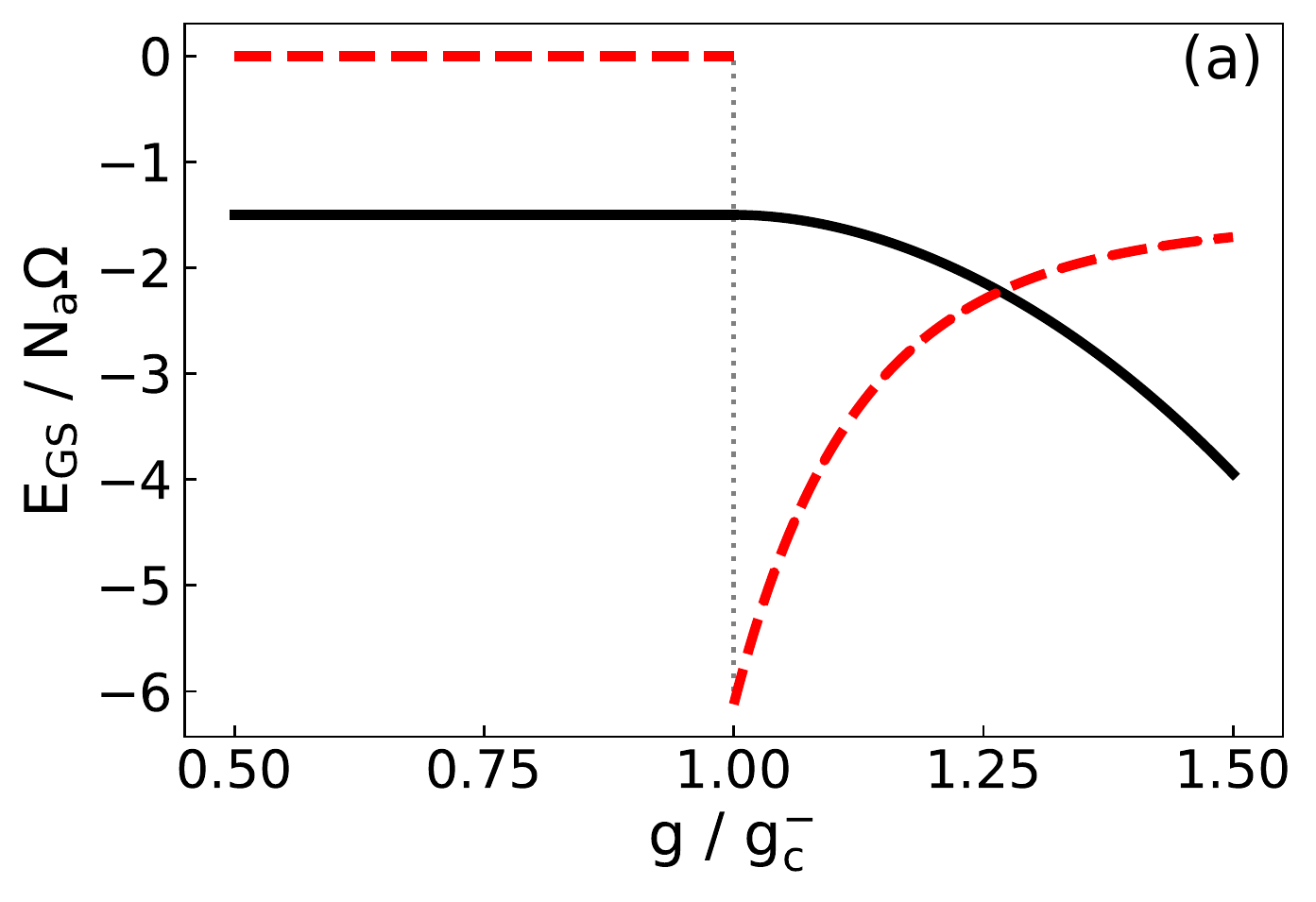} 
\includegraphics[height=0.20\linewidth]{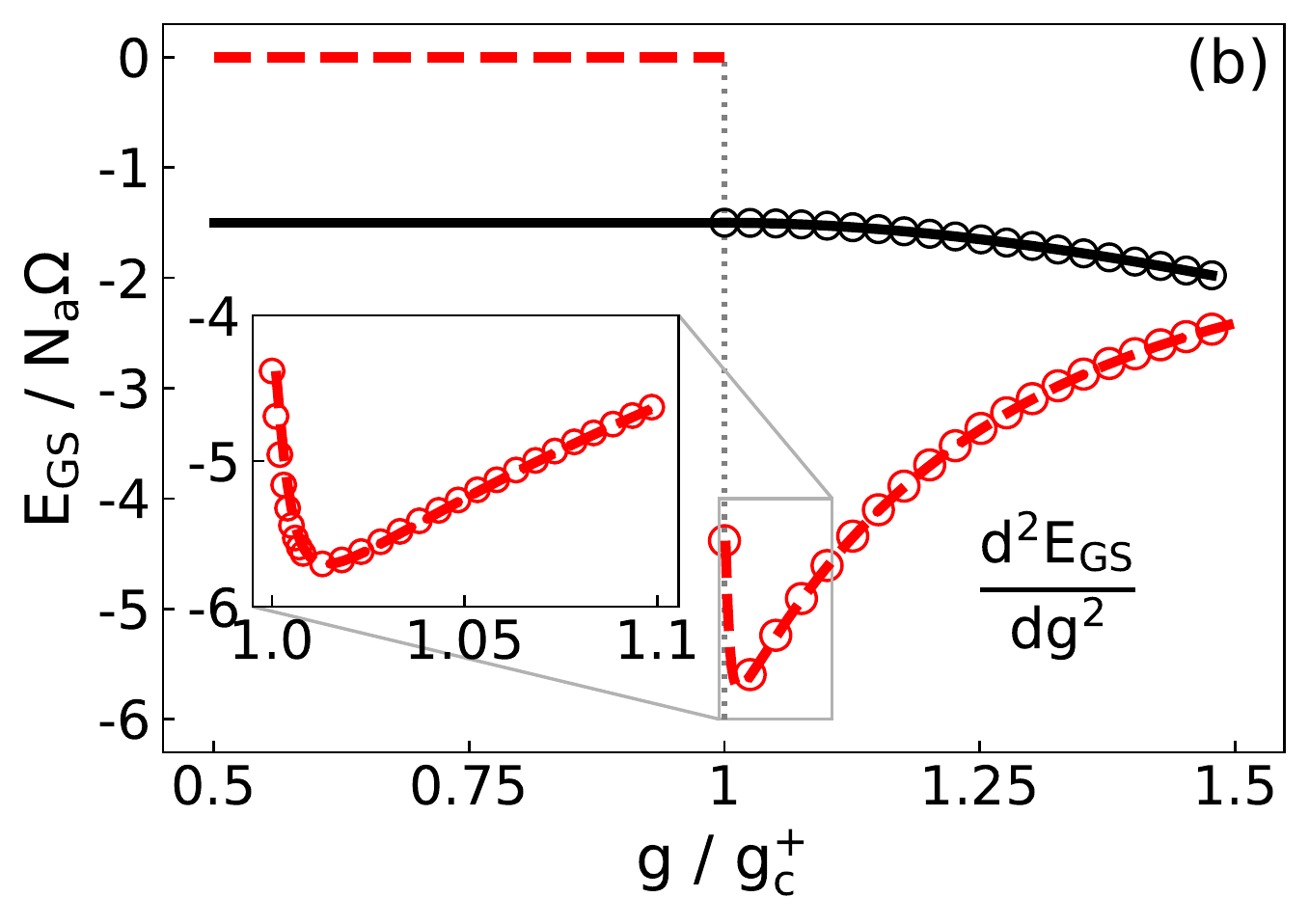} 
\includegraphics[height=0.20\linewidth]{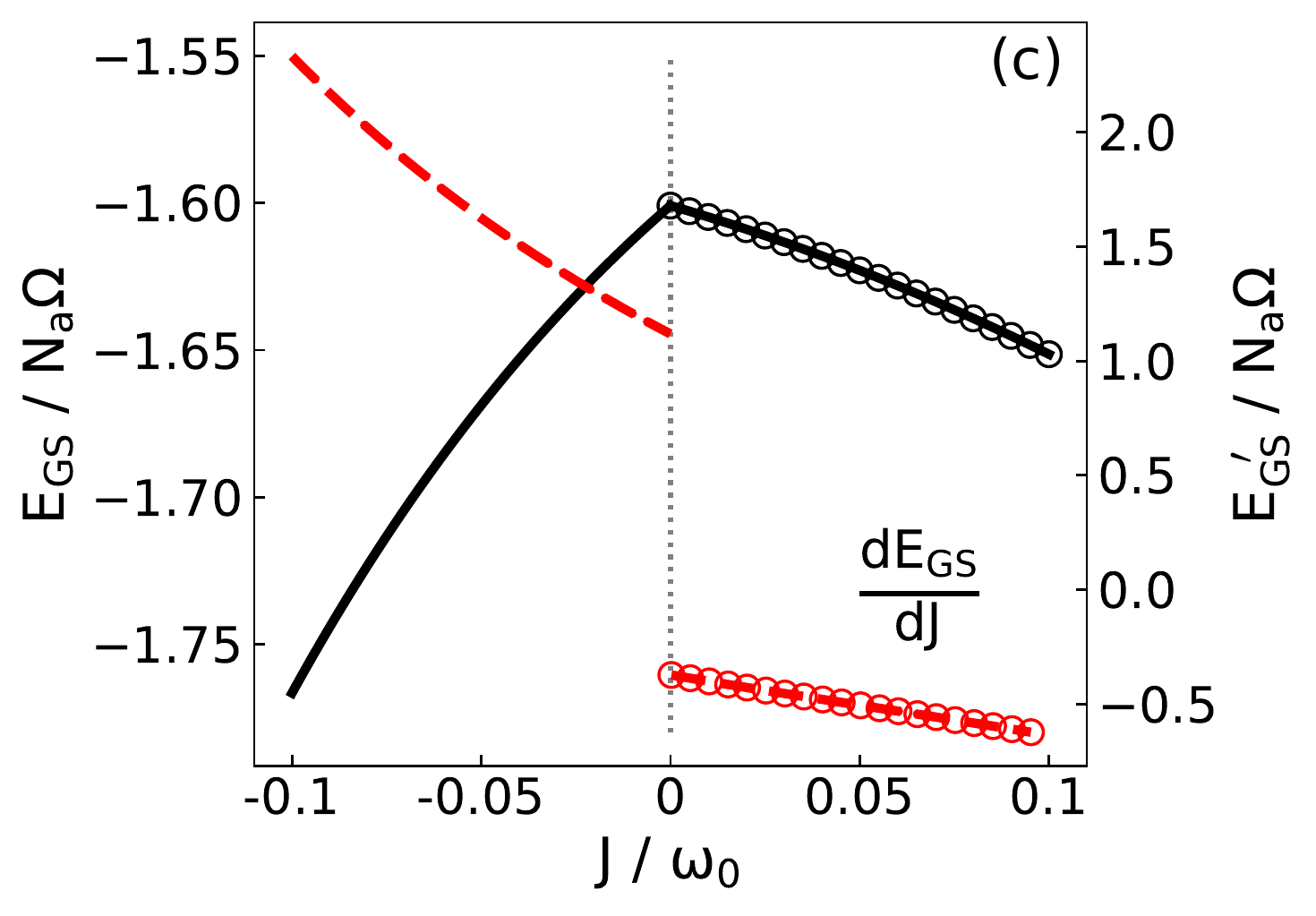} \\
\caption{(a, b) The rescaled ground-state energy (black line) and its second-order derivative (red dashed line) as a function of $g/g_c^{\mp}$ for (a) $J<0$  and for (b) $J>0$ . Circles indicate numerical solutions. Both show that the ground state energy $E_{\mathrm{GS}}$ is discontinuous in the second-order derivative in $g$. (c) The rescaled ground-state energy (black line) and its first-order derivative (red dashed line) as a function of the dimensionless hopping strength $J/\omega_0$ in the superradiant phase ($g=1.2$). The first-order derivative is discontinuous indicating a first-phase transition at $J/\omega_0=0$.}
\label{fig:gs}
\end{figure}

Finally, from the ground-state energy studied in this section, we show that both the non-frustrated and frustrated are the second-order phase transition (see Fig. \ref{fig:gs}). For $J<0$, $d^2E_{\mathrm{GS}}/dg^2$ is monotonically increasing for $g>g_c^-$, which is consistent with the behavior of QPT for the single Dicke model \cite{emary_chaos_2003_s}. Interestingly, we observe that the second derivative of the ground state energy for $J>0$ exhibits non-monotonicity where there exists a point $g_0>g_c^+$, below which it monotonically decreases and above which it monotonically increases. At $g=g_c^{+}$, we have $d^2E_{\mathrm{GS}}/dg^2=-4/(g_c^+)^2$, which can be derived from Eq. (\ref{eq:gs_sp2_exp}). Meanwhile, the ground-state energy $E_{\mathrm{GS}}$ as a function of intercavity hopping $J$ exhibits a discontinuity in the first order derivative in the superradiant phase, indicating a first-order QPT at $J=0$, where we have three decoupled Dicke model. Therefore, $g=1,J=0$ locates a tricritical point.

\section{III. \quad Excitation Spectra for the Dicke Trimer}
In this section, we will present the solution for the quadratic Hamiltonian of the Dicke trimer for both signs of $J$. To begin with, we express the quadratic Hamiltonian given in Eq. (\ref{eq:Hq}) using cavity quadratures $q_{n}=\left(a_{n}+a_{n}^{\dagger}\right) / \sqrt{2}$, $p_{n}=i\left(a_{n}^{\dagger}-a_{n}\right) / \sqrt{2}$ and atomic quadratures $Q_{n}=\left(b_{n}+b_{n}^{\dagger}\right) / \sqrt{2}$, $P_{n}=i\left(b_{n}^{\dagger}-b_{n}\right) / \sqrt{2}$,
\begin{equation}\label{eq:Hq_pq}
H_q=\sum_{n=1}^3\left[\frac{\omega_{0}}{2} \left(q_n^2+p_n^2\right)-\frac{\Omega}{2\cos{\theta_n}} \left(Q_n^2+P_n^2\right)+2\lambda\cos{\theta_n}\cos\phi_n Q_n q_n+J(q_n q_{n+1}+p_n p_{n+1})\right],
\end{equation}
which is Eq. (8) in the main text. ${H}_q$ is a bilinear Hamiltonian in terms of quadrature operators and can  therefore be written in the form 
\begin{equation}\label{eq:form}
    {H}_q=\frac{1}{2} \boldsymbol{r}^{\top} \mathcal{H}_q \boldsymbol{r},
\end{equation}
where $\boldsymbol{r}=\left(q_{1}, p_{1}, Q_{1}, P_{1},q_{2}, p_{2}, Q_{2}, P_{2},q_{3}, p_{3}, Q_{3}, P_{3}\right)^{\top}$ and $\mathcal{H}_q $ is a symmetric, positive-definite $12 \times 12$ matrix
\begin{equation}\label{eq:Ksp2}
\mathcal{H}_q=
\left(\begin{array}{ccc}
\mathcal{H}_1 & \mathcal{H}_J & \mathcal{H}_J\\
\mathcal{H}_J & \mathcal{H}_2 & \mathcal{H}_J\\
\mathcal{H}_J & \mathcal{H}_J & \mathcal{H}_3.
\end{array}\right).
\end{equation}
Here, $\mathcal{H}_n$ and $\mathcal{H}_J$ are defined by
\begin{equation}
\mathcal{H}_n=
\left(\begin{array}{cccc}
\omega_{0} & 0 & g \cos{\theta_n}\cos\phi_n  \sqrt{\omega_0  \Omega } & 0 \\
0 & \omega_{0} & 0 & 0 \\
g\cos{\theta_n}\cos\phi_n  \sqrt{\omega_0  \Omega } & 0 & -{\Omega }/{\cos{\theta_n}} & 0 \\
0 & 0 & 0 & -{\Omega }/{\cos{\theta_n}}
\end{array}\right),\quad
\mathcal{H}_J=
\left(\begin{array}{cccc}
J & 0 & 0 & 0 \\
0 & J & 0 & 0 \\
0 & 0 & 0 & 0 \\
0 & 0 & 0 & 0
\end{array}\right).
\end{equation}
According to Williamson theorem \cite{serafini_quantum_2017}, one can find a symplectic matrix $S$ such that
\begin{equation}
 S \mathcal{H}_q S^{\top}=V, \quad V=\operatorname{diag}\left(\varepsilon_{1}, \varepsilon_{1}, \varepsilon_{2}, \varepsilon_{2}, \varepsilon_{3}, \varepsilon_{3}, \varepsilon_{4}, \varepsilon_{4}, \varepsilon_{5}, \varepsilon_{5}, \varepsilon_{6}, \varepsilon_{6}\right)
\end{equation}
where $\varepsilon_{n} \geq 0$ are the symplectic eigenvalues computed by taking modulus of the 12 eigenvalues of the matrix $i \Omega_0 \mathcal{H}_q$. A symplectic matrix satisfies $S \Omega_0 S^{\top}=\Omega_0$, where the symplectic form, $\Omega_0$ is defined by
\begin{equation}
\Omega_0=\bigoplus_{n=1}^{6}\left(\begin{array}{cc}
0 & 1 \\
-1 & 0
\end{array}\right).
\end{equation}
Transforming to a new set of quadratures $\boldsymbol{r}^{\prime}=(S^{-1})^\top \boldsymbol{r}=\left(q_{1}^{\prime}, p_{1}^{\prime}, q_{2}^{\prime}, p_{2}^{\prime}, q_{3}^{\prime}, p_{3}^{\prime}, q_{4}^{\prime}, p_{4}^{\prime},  q_{5}^{\prime}, p_{5}^{\prime},  q_{6}^{\prime}, p_{6}^{\prime}\right)^{\top}$ diagonalizes $H_q$ as
\begin{equation}
	H_q=\frac{1}{2} \sum_{n=1}^{6} \varepsilon_{n}\left(q_{n}^{\prime 2}+p_{n}^{\prime 2}\right).
\end{equation}
The ground state is Gaussian and therefore can be characterized by the covariance matrix \cite{serafini_quantum_2017}
\begin{equation}\label{eq:covariance}
	\mathcal{C}=\frac{1}{2}S^{\top}S.
\end{equation}
In our model, the covariance matrix need to be computed numerically, from which we obtain the cavity population and squeezing variances.

\subsection{A. \quad Normal Phase}
At the origin, $\bar\alpha_{1,2,3}=0$, Eq. (\ref{eq:Hq_pq}) becomes
\begin{equation}\label{eq:npeff}
{H}_{\mathrm{np}}= \sum_{n=1}^3\left[\frac{\omega_{0}}{2} \left(q_n^2+p_n^2\right)+\frac{\Omega}{2} \left(Q_n^2+P_n^2\right)-g\sqrt{\omega_0\Omega} Q_n q_n+J(q_n q_{n+1}+p_n p_{n+1})\right],
 \end{equation}
which respects the translational symmetry. We then perform a Fourier transformation $q_{n}=\sum_{k} e^{-i k n} q_{k} / \sqrt{3}$, $Q_{n}=\sum_{k} e^{-i k n} Q_{k} / \sqrt{3}$ with quasimomentum $k=0, \pm 2 \pi / 3$ to get
\begin{equation}
	H_{\mathrm{np}}= \sum_{k}\left[\frac{\omega_{0}}{2} \left(q_kq_{-k}+p_kp_{-k}\right)+\frac{\Omega}{2} \left(Q_kQ_{-k}+P_kP_{-k}\right)-g\sqrt{\omega_0\Omega} Q_k q_{-k}+J\left(e^{ik}q_k q_{-k}+e^{-ik}p_k p_{-k}\right)\right].
\end{equation}
It can be observed that the zero momentum mode $k=0$ is decoupled from the other two modes $k=\pm2/3\pi$, so we diagonalize the zero momentum mode and finite momentum modes separately by symplectic transform. {The excitation of the zero momentum mode is}
\begin{equation}
 \varepsilon_{0(\pm)}^{\mathrm{np}} = \frac{\omega_0}{\sqrt2}\left\{(1+2\bar J)^2+\bar\Omega^2\pm\sqrt{\left[(1+2\bar J)^2-\bar\Omega^2\right]^2+4\bar\Omega^2(1+2\bar J)g^2}\right\}^{1/2},
\end{equation}
where $\bar\Omega=\Omega/\omega_0$. Two finite momentum modes ($k=\pm 2\pi/3$) have degenerate excitation energies,
\begin{equation}
  \varepsilon_{+\frac{2\pi}{3}(\pm)}^{\mathrm{np}} = \varepsilon_{-\frac{2\pi}{3}(\pm)}^{\mathrm{np}} = \frac{\omega_0}{\sqrt2}\left\{(1-\bar J)^2+\bar\Omega^2\pm\sqrt{\left[(1-\bar J)^2-\bar\Omega^2\right]^2+4\bar\Omega^2(1-\bar J)g^2}\right\}^{1/2}.
\end{equation}
Here, $(-)$ denotes the lower branch in the energy spectra, while $(+)$ denotes the higher branch. Only the lower branch of the energy spectra is shown in Fig. 2 of the main text. Crucially, the excitation energy $\varepsilon_{0(-)}^{\mathrm{np}}$ is real only when
\begin{equation}\label{eq:gc<}
    (1+2\bar J)^2+\bar\Omega^2>
    \sqrt{\left[(1+2\bar J)^2-\bar\Omega^2\right]^2+4\bar\Omega^2(1+2\bar J)g^2},
\end{equation}
or equivalently, $g<g_c^-$.  On the other hand, the excitation energies $\varepsilon_{\pm\frac{2\pi}{3}(-)}^{\mathrm{np}}$ are real only when
\begin{equation}\label{eq:gc>}
    (1-\bar J)^2+\bar\Omega^2>\sqrt{\left[(1-\bar J)^2-\bar\Omega^2\right]^2+4\bar\Omega^2(1-\bar J)g^2},
\end{equation}
or equivalently, $g<g_c^+$. Therefore, we conclude that i) for $J<0$, the zero momentum mode becomes critical and ii) for $J>0$, the two degenerate finite momentum modes become critical. Moreover, in both cases, the critical excitation energy closes their energy gap with an exponent $1/2$, namely, 
\begin{eqnarray}
	  {\varepsilon_{0(-)}^{\mathrm{np}}} (g\sim g_c^-)&\propto&|g-g_c|^{1/2}\quad (J<0),\nonumber\\
	  {\varepsilon_{\pm\frac{2\pi}{3}(-)}^{\mathrm{np}}} (g\sim g_c^+)&\propto&|g-g_c|^{1/2}\quad (J>0).
\end{eqnarray}

\subsection{B. \quad Non-frustrated Superradiant Phase ($J<0$)}

\begin{figure}[t]
\centering
\includegraphics[width=0.3\textwidth]{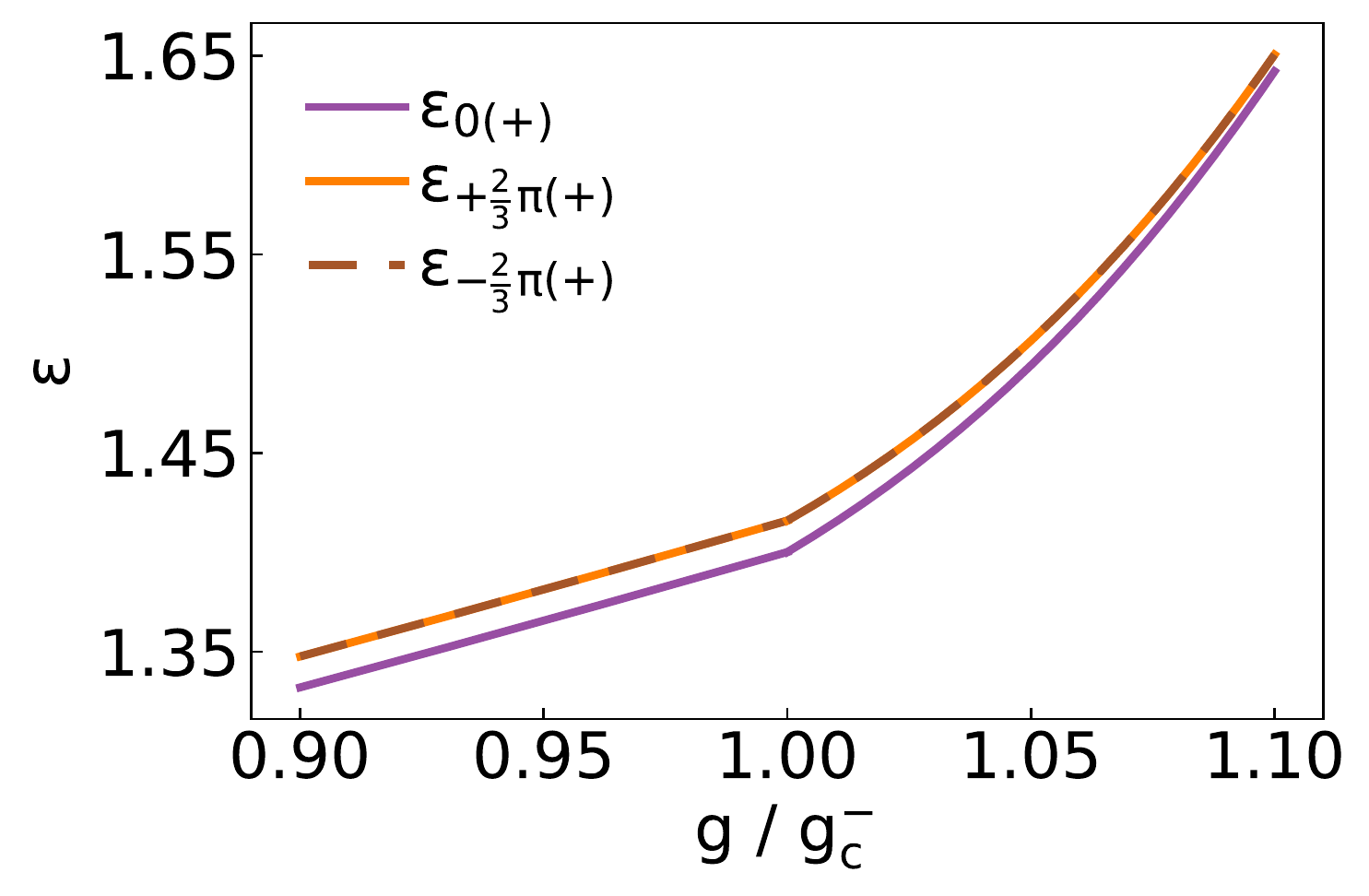}
\caption{Excitation spectra in the higher branch (+) as a function of the dimensionless coupling strength $g/g_c^{-}$ for $J<0$.}
\label{fig:nfsphigh}
\end{figure}

For $J<0$ and $g>g_c^-$, we insert the non-frustrated mean-value solutions, given in Eq. (7) from the main text, to Eq. (\ref{eq:Hq_pq}) and derive the effective Hamiltonian in the non-frustrated superradiant phase (NFSP)
\begin{equation}
	{H}_{\mathrm{nfsp}}=\sum_{n=1}^3\left[\frac{\omega_{0}}{2} \left(q_n^2+p_n^2\right)+\frac{\Omega'}{2} \left(Q_n^2+P_n^2\right)-g'\sqrt{\omega_0\Omega} Q_n q_n+J(q_n q_{n+1}+p_n p_{n+1})\right],
\end{equation}
where $\Omega^{\prime}=\left({g}/{g_c^-}\right)^2\Omega$ and $g^{\prime}=(g_c^-)^2/g$. Since the translational symmetry is preserved, we perform a Fourier transformation as done in the normal phase to find
\begin{equation}
	{H}_{\mathrm{nfsp}}= \sum_{k}\left[\frac{\omega_{0}}{2} \left(q_kq_{-k}+p_kp_{-k}\right)+\frac{\Omega'}{2} \left(Q_kQ_{-k}+P_kP_{-k}\right)-g'\sqrt{\omega_0\Omega} Q_k q_{-k}+J\left(e^{ik}q_k q_{-k}+e^{-ik}p_k p_{-k}\right)\right].
\end{equation}
Note that the zero momentum mode is decoupled from finite momentum modes. Therefore, we first derive the excitation energies of the zero momentum mode as
{
\begin{equation}
  {\varepsilon_{0(\pm)}^{\mathrm{nfsp}}} = \frac{\omega_0}{\sqrt2}\left\{\frac{g^4\bar\Omega^2}{(1+2\bar J)^2}+(1+2\bar J)^2\pm\sqrt{\left[\frac{g^4\bar\Omega^2}{(1+2\bar J)^2}-(1+2\bar J)^2\right]^2+4\bar\Omega^2(1+2\bar J)^2}\right\}^{1/2}, \\
\end{equation}
}
and the excitation energies of finite momentum modes are given by
 \begin{equation}
  {\varepsilon_{+\frac{2\pi}{3}(\pm)}^\mathrm{nfsp}} = {\varepsilon_{-\frac{2\pi}{3}(\pm)}^{\mathrm{nfsp}}} = \frac{\omega_0}{\sqrt2}\left\{\frac{g^4\bar\Omega^2}{(1+2\bar J)^2}+(1-\bar J)^2\pm\sqrt{\left[\frac{g^4\bar\Omega^2}{(1+2\bar J)^2}-(1-\bar J)^2\right]^2+4\bar\Omega^2(1+2\bar J)(1-\bar J)}\right\}^{1/2}
\end{equation}
The excitation energy $\varepsilon_{0(-)}^{\mathrm{nfsp}}$ is real only when
\begin{equation}
\frac{g^4\bar\Omega^2}{(1+2\bar J)^2}+(1+2\bar J)^2>\sqrt{\left[\frac{g^4\bar\Omega^2}{(1+2\bar J)^2}-(1+2\bar J)^2\right]^2+4\bar\Omega^2(1+2\bar J)^2},
\end{equation}
or equivalently, $g>g_c^-$. Therefore, for $J<0$, it is still the zero momentum mode that becomes critical in the superradiant phase. Moreover, near the critical point, the zero momentum mode closes its energy gap with an exponent $1/2$,
\begin{equation}
	\varepsilon_{0(-)}^{\mathrm{nfsp}} (g\sim g_c^-)\propto|g-g_c|^{1/2}\quad (J<0)
\end{equation}
The excitation energies in the lower branch (-) are shown in Fig. 2(a) from the main text, and that of the higher branch (+) are shown in Fig. (\ref{fig:nfsphigh}).

\subsection{C. \quad Frustrated Supperradiant Phase ($J>0$)}

\begin{figure}[t]
\centering
{{\includegraphics[width=0.3\linewidth]{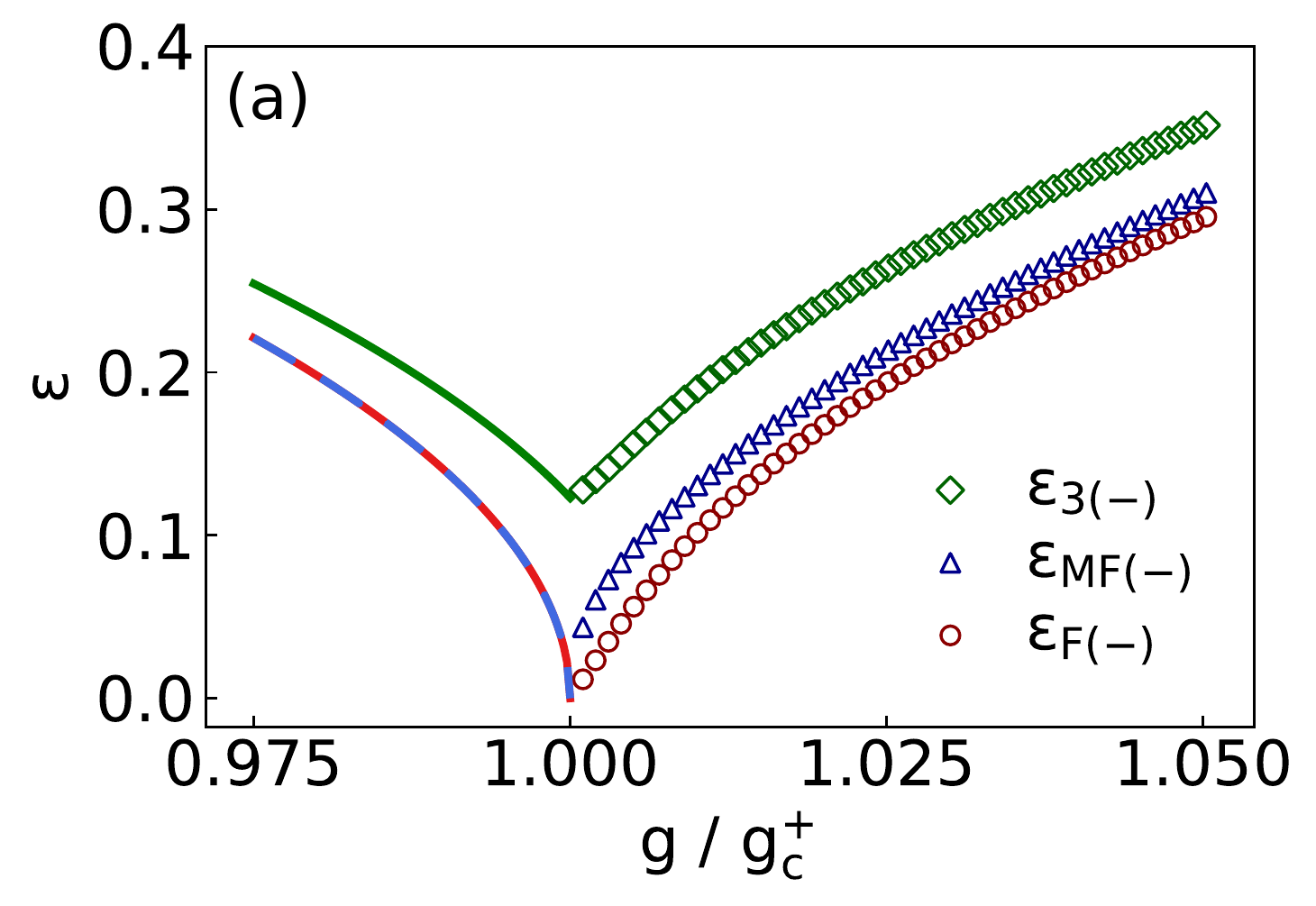} }}
{{\includegraphics[width=0.3\linewidth]{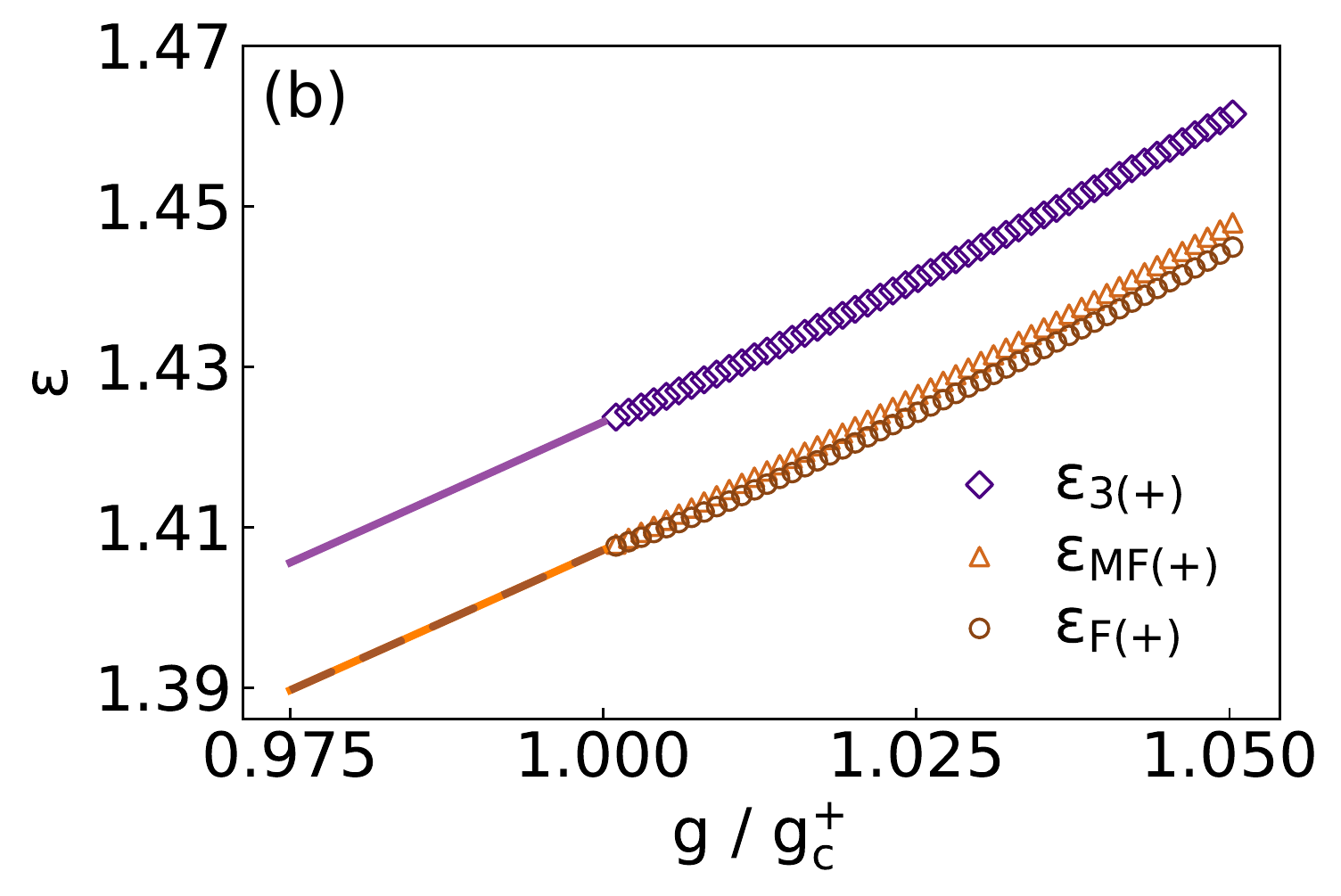} }}
\caption{Excitation spectra of the Dicke trimer in (a) the lower branch and (b) the higher branch for $J>0$. Note that here the range of the coupling strength $g/g_c^{+}\in (0.975,1.05)$ is larger than the one in Fig. 2 (b) in the main text. Lines represent analytical results and shapes represent numerical results. In both branches, the degeneracy of the normal modes are lifted in the frustrated superradiant phase. In the lower branch, $\varepsilon_{\mathrm{F}}$ and $\varepsilon_{\mathrm{MF}}$ get closer when $|g-g_c^+|$ gets larger.}
\label{fig:excitationsp22}
\end{figure}

When $J>0$, the mean-value solutions given in Eq. (6) of the main text and Eq. (\ref{eq:Ftheta}) break the translational symmetry and so does the resulting quadratic Hamiltonian. We first  diagonalize Eq. (\ref{eq:Hq_pq}) directly by a symplectic transform, and then use the approximate solution in Eq. (\ref{eq:Ftheta}) to investigate critical behaviors in the excitation spectra. We find that there are two critical excitation energies, and the components of corresponding normal modes are provided in Fig. 2(c) and (d) from the main text. There exists a mean-field mode where contributions of the cavities $2$ and $3$ are equal and have an opposite sign with that of the cavity $1$. For the corresponding atomic quadratures, we note that $\phi_1=0$ and $\phi_2=\phi_3=\pi$ for the third term in Eq. (\ref{eq:Hq_pq}), so $Q_1$ has the same sign with $Q_2,Q_3$ in Fig. 2(d). Meanwhile, there is a frustrated mode that consists only of the cavities $2$ and $3$ and decouples the cavity 1, whose excitation energy reads 
\begin{equation}
	\varepsilon_{\mathrm{F}}^2 = \frac{\omega_0^2}{2}\left\{(g_c^+)^4+\left(\frac{\bar\Omega}{\cos{\theta_2}}\right)^2-\sqrt{\left[(g_c^+)^4-\left(\frac{\bar\Omega}{\cos{\theta_2}}\right)^2\right]^2+4\bar\Omega^2(g_c^+)^2g^2\cos{\theta_2}}\right\}.
\end{equation}
Note that $\cos{\theta_2}$ is a function of $g$ as given in Eq.~(\ref{eq:Ftheta}). We then expand $\varepsilon_{\mathrm{F}}^2$ around $g=g_c^{+}$ and plug in $\cos{\theta_2}(g_c^+)=-1$ to get
\begin{equation}\label{eq:softexpand}
 \varepsilon_{\mathrm{F}}^2 = \frac{2\omega_0^2\bar\Omega^2(g_c^+)^4}{(g_c^+)^4+\bar\Omega^2}\left[3\cos{\theta_2}^{\prime}|_{g=g_c^+}-\frac{2}{g_c^+}\right](g-g_c^+)+O\left((g-g_c^+)^2\right).
\end{equation}
where $\cos{\theta_2}^{\prime}(g)$ denotes the {first} derivative with respect to $g$. From Eq. (\ref{eq:Ftheta}a) we have $\cos{\theta_2}^{\prime}(g_c^+)=\frac{2}{3g_c^+}$, so that the first order term of Eq.~ (\ref{eq:softexpand}) vanishes. Note that $\cos{\theta_2}^{\prime}(g)$ is determined by the first term in $\bar\alpha_2$ from Eq.(6b) in the main text. To derive the second order term in Eq. (\ref{eq:softexpand}), one has to take into account the second order term in $\cos\theta_2$, or equivalently, the second term in $\bar\alpha_2$. In short, we show that
\begin{equation}
	\varepsilon_{\mathrm{F}} = 2\omega_0\sqrt{\frac{(1-\bar J)(2+\bar J)}{3\bar J\left[(1-\bar J)^2+\bar\Omega^2\right]}}\left(g-g_c^+\right)+O\left(\left(g-g_c^+\right)^2\right).
\end{equation}
which is the expression we use for Fig. 2(b) in the main text. On the other hand, we find that the lowest order term of the mean-field mode is given by
\begin{equation}
 \varepsilon_{\mathrm{MF}} = 2\omega_0\sqrt{\frac{(1-\bar J)^{3/2}}{(1-\bar J)^2+\bar\Omega^2}}\left(g-g_c^+\right)^{1/2}+O\left(\left(g-g_c^+\right)^{3/2}\right).
\end{equation}
We note that while the critical behavior of the frustrated and mean-field mode near critical point is qualitatively different with each other with distinct critical exponents, they become nearly degenerate for as $g$ moves away from the critical point, as shown in Fig. \ref{fig:excitationsp22}(a).

The non-mean-field exponent $1$ for the frustrated mode and the mean-field exponent $1/2$ for the MF mode can be understood from the curvature of the mean-field energy functions near the frustrated superradiant solution, which is captured by the eigenvalues of Hessian given in Eq. (\ref{eq:beta}). By inserting Eq. (6) from the main text to $\beta_1$ from Eq. (\ref{eq:beta}), which is a general form for the eigenvalue of Eq. (\ref{eq:hessian}) when $\bar\alpha_2=\bar\alpha_3$, and expanding around $g=g_c^+$, one finds
\begin{equation}
	\beta_{\mathrm{F}}\simeq\frac{4 (2+\bar J) }{3 \bar J}\left(g-g_c^+\right)^2+O\left(\left(g-g_c^+\right)^3\right).
\end{equation}
with an eigenvector $\textbf{y}_\textrm{F}=(0,1,-1)$. On the other hand, by inserting Eq. (6) to $\beta_2$ from Eq. (\ref{eq:beta}) and expanding around $g=g_c^+$, one finds
\begin{equation}\label{eq:oddpd}
	\beta_{\mathrm{MF}}\simeq8 \sqrt{1-\bar J} \left(g-g_c^+\right)+O\left(\left(g-g_c^+\right)^2\right),
\end{equation}
whose corresponding eigenvector can be written as $\textbf{y}_\textrm{MF}=(y_1,y_2,y_2)$, where $y_2<0<y_1$. Note that $\beta_{\mathrm{MF}}$ is typical behavior of Landau potential when expanded around the minimum of a double well potential and it gives rise to the closing gap with a mean-field exponent because $\varepsilon_\textrm{MF}\propto\sqrt{\beta_{\mathrm{MF}}}$. In the Landau potential, however, all transverse directions to the mean-field vector $\textbf{y}_\textrm{MF}$ have a finite curvature, making all the other modes remain gapped at the critical point. The essential non-mean-field behavior of the frustrated superradiance therefore stems from the emergence of a transverse mode $\textbf{y}_\textrm{F}$, that is orthogonal to $\textbf{y}_\textrm{MF}$, whose curvature at the critical point vanishes slower than the mean-field direction, leading to a soft mode $\varepsilon_F$ with a non-mean-field exponent.
\begin{figure}[t]
\centering
{{\includegraphics[height=0.20\textwidth]{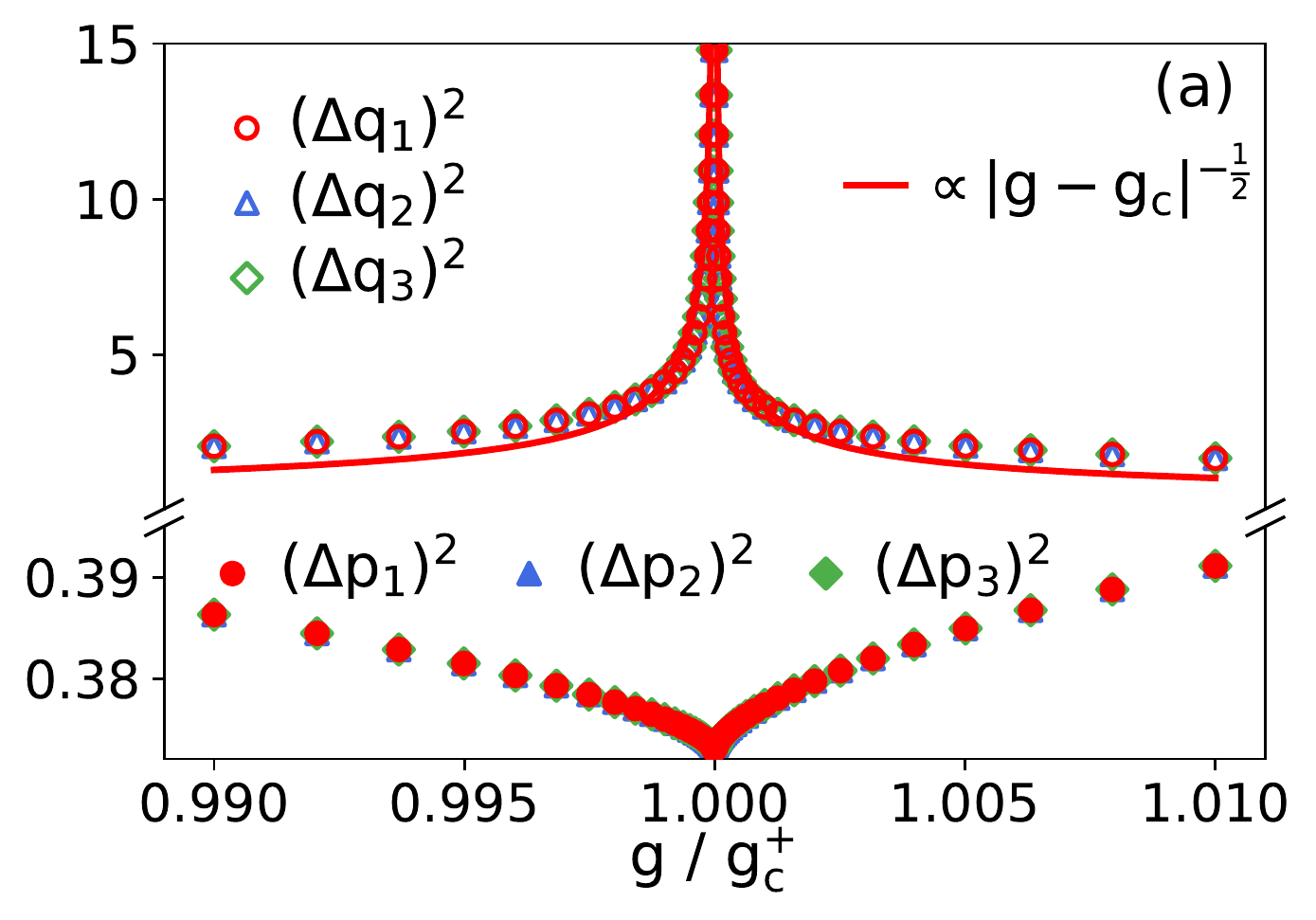} }}\quad
{{\includegraphics[height=0.20\textwidth]{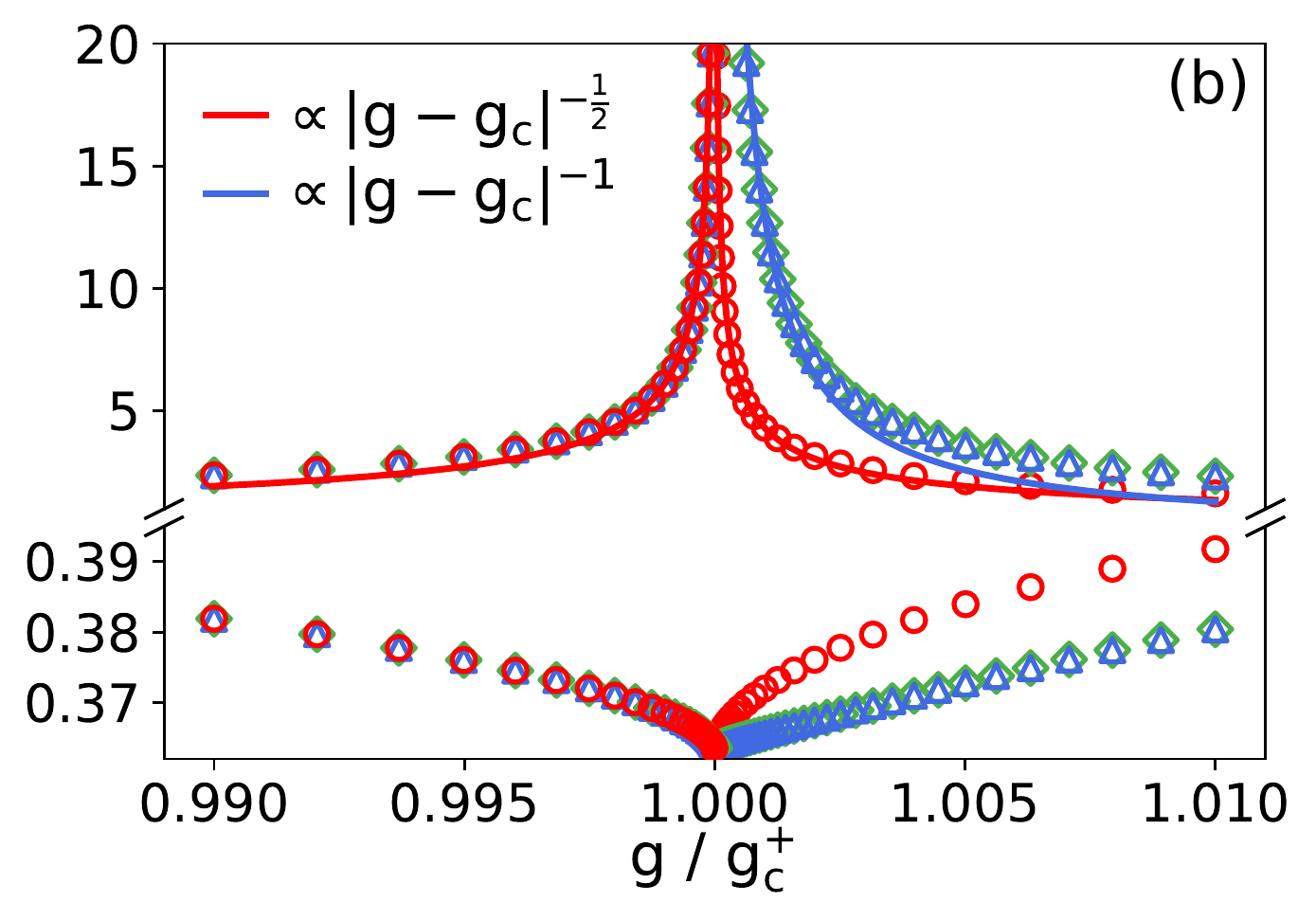} }}\quad
\caption{Squeezing {of the cavity fields in} the ground state {as a function of} $g/g_c^{\pm}$ for (a) $J<0$ and (b) $J>0$. In the non-frustrated superradiant phase, {quadrature} variances of the three cavities are identical and have the critical exponent {$1/2$}. In the frustrated superradiant phase, the translational symmetry is broken and two critical exponents $\gamma_\textrm{MF}=1/2$ and $\gamma_\textrm{F}=1$ coexist.}
\label{fig:squeezing}
\end{figure}

\section{IV. \quad {Squeezing} for the Dicke Trimer}

The squeezing variances can be numerically derived from the covariance matrix from Eq. (\ref{eq:covariance}). As shown in Fig.~\ref{fig:squeezing}, we find that in the normal phase and non-frustrated superradiant phase, all cavities have $(\Delta q_n)^2 \propto |g-g_c^{\pm}|^{-\gamma_\textrm{MF}}$. However, in the frustrated superradiant phase, we find $(\Delta q_{1})^2 \propto |g-g_c^+|^{-\gamma_\textrm{MF}}$ and $(\Delta q_2)^2 =(\Delta q_3)^2 \propto |g-g_c^{\pm}|^{{-\gamma_\textrm{F}}}$ . Note that the first cavity is displaced by $\alpha_1$ which is anti-aligned with $\alpha_2=\alpha_3$.

\section{V. \quad Generalization to Odd Number of Sites}
\begin{figure}[b]
\centering
{\includegraphics[height=0.25\textwidth]{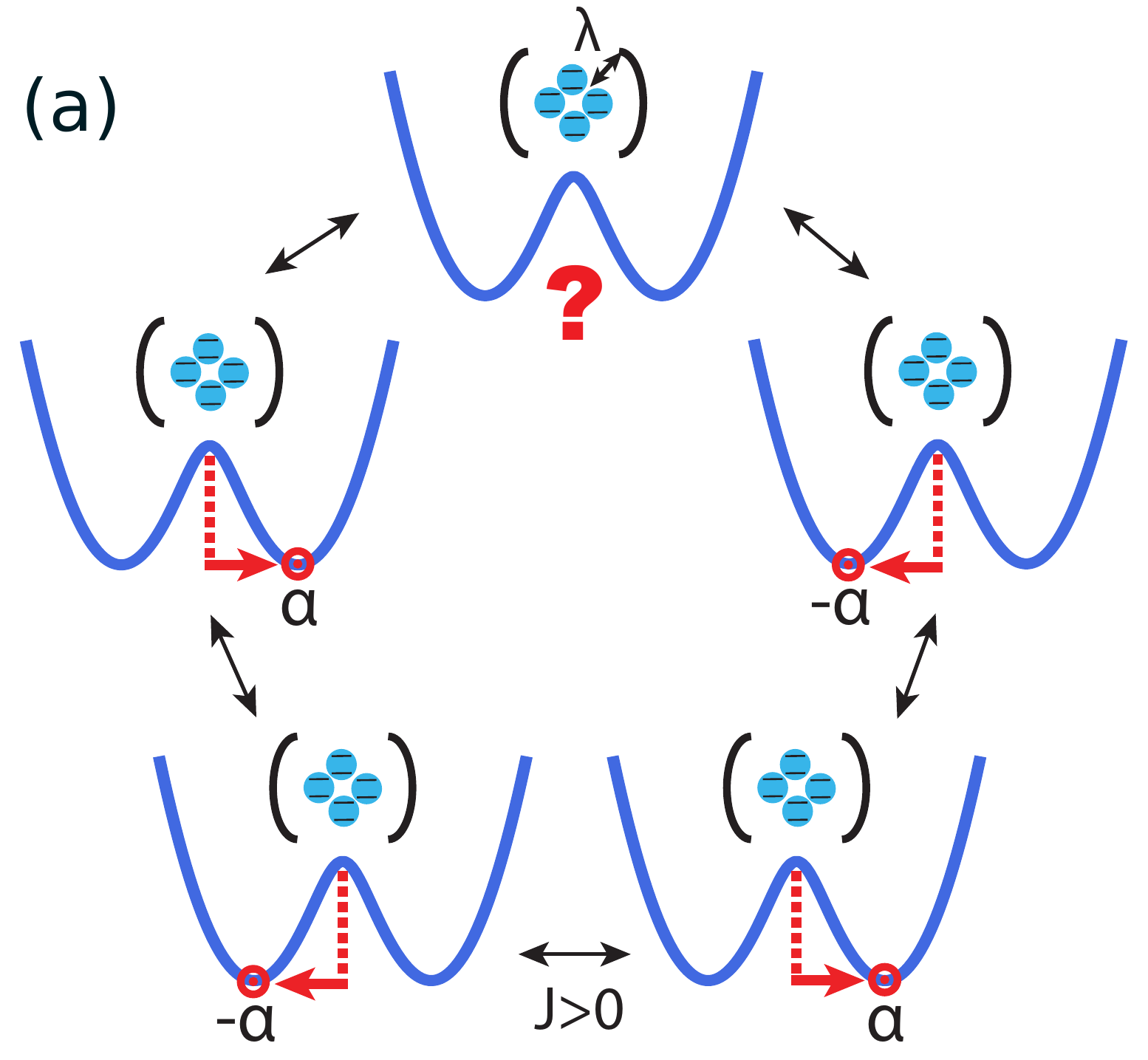} }
%{\includegraphics[height=0.24\textwidth]{schematic_5(b)} }
{\includegraphics[height=0.2\textwidth]{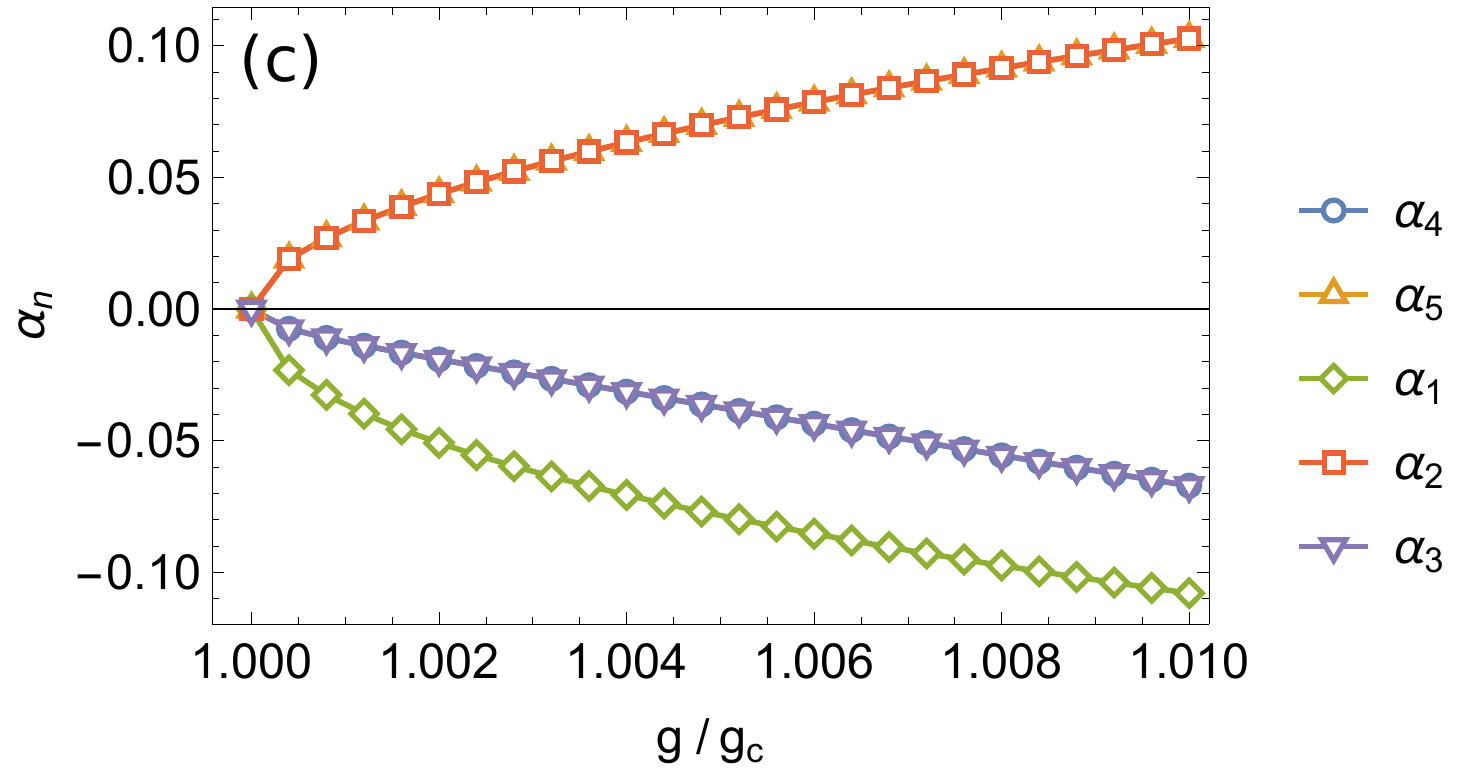} }
\caption{(a) Illustration for the mechanism of frustration for $N=5$ when neighboring $\alpha_n$ being anti-aligned becomes incompatible with the underlying geometry. Similarly, frustration will occur for any odd $N$. (b) Frustrated superradiance gives rise to a single unpaired cavity (cavity 1) and (N-1)/2 aligned pairs that respect a mirror symmetry about the unpaired cavity. (c) Numerical solution of $\bar\alpha_n(g)$ for $N=5$, where $\bar\alpha_2=\bar\alpha_5$, $\bar\alpha_3=\bar\alpha_4$ and $\bar\alpha_1$ is isolated as schematically depicted in (b). Here we show one configuration of $\alpha_n$, and there are $2N$ degenerate ground states in total.}
\label{fig:schematic5}
\end{figure}

Here, we consider the Dicke lattice model with an odd $N$ number of sites. As shown in Fig. \ref{fig:schematic5} (a), given the positive hopping energies, an odd number of sites may give rise to a frustrated superradiant phase. Solving this model requires solving $N$ coupled linear equations and diagonalizing $4N\times4N$ matrices, which are in general analytically intractable. Nonetheless, here by exploiting appropriate symmetries of the model we obtain important analytical results. In Sec. A, we use the translational symmetry of the system in the normal phase to derive an analytical expression of the critical coupling $g_c^+(N)$. In Sec. B, we use the mirror symmetry in the frustrated superradiant phase, which is caused by the pairs of identical modes described in the main text and Fig. \ref{fig:schematic5} (b) and (c) {to} show that the frustrated mode always does not involve the unpaired cavity.

\subsection{A. \quad Analytic expression for the critical point $g_c^+(N)$}
The Hessian matrix of Eq.(4) in the main text is given by
\begin{equation}\label{eq:oddhessian}
	\textrm{Hess}_{N}(\{\bar\alpha_n\})=
\left(\begin{array}{ccccc}\bar 
2-\frac{2 g^2}{\left(1+4g^2\bar{\alpha}_1^2\right)^{3/2}} & 2\bar J & 0  & \ldots & 2\bar J \\
2\bar J & 2-\frac{2 g^2}{\left(1+4g^2\bar{\alpha}_2^2\right)^{3/2}} & 2\bar J   & \ldots & 0 \\
0 & 2\bar J & \ddots & \ddots & 0 \\
\vdots & \vdots & \ddots & \ddots & 2\bar J \\
2\bar J & 0 & \ldots & 2\bar J & 2-\frac{2 g^2}{\left(1+4g^2\bar{\alpha}_{N}^2\right)^{3/2}}
\end{array}\right).
\end{equation}
Therefore, the Hessian matrix at origin is given by
\begin{equation}
\textrm{Hess}_{N}(\{\bar\alpha_n=0\})=
\left(\begin{array}{cccccc}
2-2g^2 & 2\bar J & 0 & 0 & \ldots & 2\bar J \\
2\bar J & 2-2g^2 & 2\bar J & 0 & \ldots & 0 \\
0 & 2\bar J & 2-2g^2 & 2\bar J & \ldots & 0 \\
\vdots & \vdots & \vdots & \ddots & \ddots & 2\bar J \\
2\bar J & 0 & \ldots & 0 & 2\bar J & 2-2g^2
\end{array}\right){.}
\end{equation}
The critical point $g_c$ can be determined by solving $\min\{\lambda_t\}=0$, where $\{\lambda_{t=0,1,...,N-1}\}$ are the eigenvalues of $\textrm{Hess}_{N}(\{\bar\alpha_n=0\})$. We note that due to the translational symmetry in the normal phase, $\textrm{Hess}_{N}(\bar\alpha_n=0)$ is a circulant matrix. Using the method given in Ref.~\cite{golub_matrix_2013}, we find that 
\begin{equation}
	\lambda_t=1-g^2+2\bar J \cos \left(\frac{2\pi t}{N}\right) \quad \text{for} \quad t=0,1,2,\cdots,N-1{.}
\end{equation}
From the above equation, one can see the difference between the positive and negative intercavity hopping. For $J<0$, one eigenvalue, $\lambda_{t=0}$, becomes negative if $g>g_c^{-}=\sqrt{1-2\bar J}$. Therefore, the critical point becomes independent of $N$ for $J<0$. This is because it is the zero momentum mode that becomes critical, which has no spatial variation. However, for $J>0$, two eigenvalues, $\lambda_{t=(N-1)/2}$ and $\lambda_{t=(N+1)/2}$, become negative simultaneously if
\begin{equation}
	g>g_c^{+}(N)=\sqrt{1+2\bar J \cos\left(\frac{N-1}{N}\pi\right)}.
\end{equation}
The two fold degeneracy of the critical modes in $J>0$ is because they have a finite momentum, which always comes in degenerate pairs ($+$ and $-$ signs) due to the time-reversal symmetry. 

\subsection{B. \quad The frustrated mode}

For the Dicke trimer case, we have identified that the frustrated superradiant phase is characterized by the emergence of two critical scalings and that the frustrated mode decouples a single site that does not form a pair through mirror symmetry. See Fig. 2(c) and (d) from the main text. Here, we show that these two properties are generic features of the frustrated superradiant phase in a $1$D lattice. 

\begin{figure}[b]
	\includegraphics[width=0.35\textwidth]{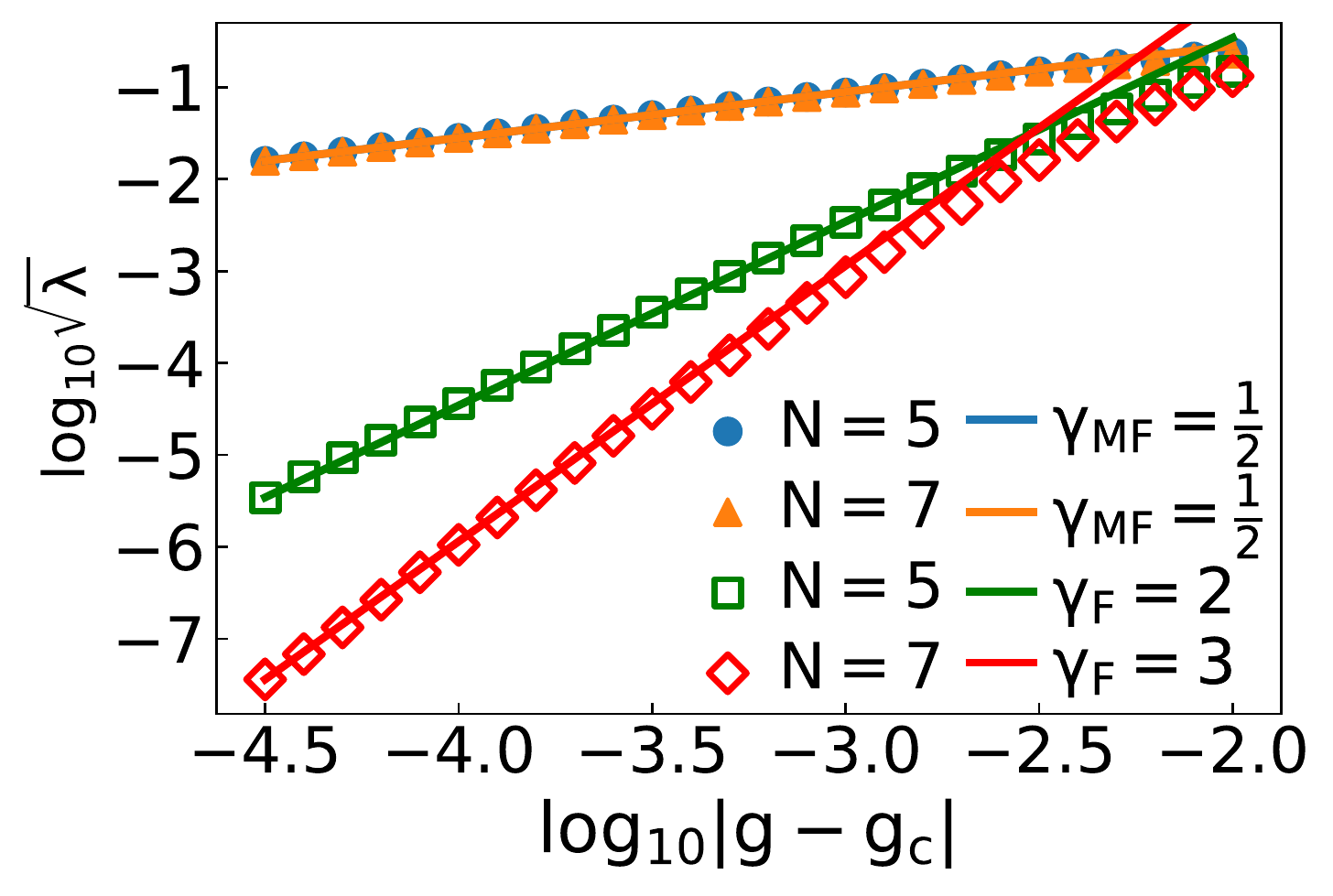}
	\caption{Scaling of the critical eigenvalues of $\mathrm{Hess}_{N}$ for $N=5,7$. Shapes are numerical solutions and lines indicate corresponding power-laws. The scaling properties agree with the results by symplectic transform, as shown in Fig. (4) of the main text.}
	\label{fig:lambdaN1}
\end{figure}

As described in the main text, we find that there are always two critical modes in $J>0$ case. As in the trimer case, we examine the scaling of the smallest eigenvalue of Eq. (\ref{eq:oddhessian}) and find $\lambda_{\mathrm{MF}}\propto|g-g_c(N)|^{1}$, $\lambda_{\mathrm{F}}\propto|g-g_c(N)|^{N-1}$ for $N=3,5,7$. The scaling of the frustrated curvature is shown in FIG. \ref{fig:lambdaN1}, which agrees with the critical scaling of the frustrated mode presented in the main text obtained by using symplectic diagonalization of the quadratic Hamiltonian. We also calculate the eigenvectors corresponding to $\lambda_{\mathrm{MF}}$ and $\lambda_{\mathrm{F}}$ and find that they have the same structure with $N=3$ case. Namely, the frustrated mode eigenvector decouples the unpaired site $1$ and has anti-symmetric weights for each ferromagnetic pair and it is always orthogonal to the mean-field mode eigenvector. Therefore, we conclude that the emergence of the frustrated mode that is orthogonal to the mean-field mode is a generic property of the frustrated superradiance of the $1$D Dicke lattice. To support this claim for any $N$, below we show that there always exists $(N-1)/2$ eigenvectors of the Hessian matrix which share the property of the frustrated mode for odd $N$, one of which we expect to become critical, given the configuration of $\alpha_n$ described in the main text and shown in Fig.~\ref{fig:schematic5} for $N=5$ case, i.e. $\alpha_{1+j}=\alpha_{N+1-j}$ where $j=1,2,...,(N-1)/2$.

\begin{proof}
Under the assumption $\alpha_{1-j}=\alpha_{N+1-j}$, we can rewrite the Hessian of Eq. (\ref{eq:oddhessian}) as
\begin{equation}\label{eq:symhessian}
\mathrm{Hess}_{N}=
\left(\begin{array}{ccccccccc}
A_1 & \bar J & 0 & 0 & 0 & 0 & 0 & 0 & \bar J \\
\bar J & A_2 & \bar J & 0 & 0 & 0 & 0 & 0 & 0 \\
0 & \bar J & A_3 & \bar J & 0 & 0 & 0 & 0 & 0 \\
0 & 0 & \bar J & \ddots & \ddots & 0 & 0 & 0 & 0 \\
0 & 0 & 0 & \ddots & A_{(N+1)/2} & \bar J & 0 & 0 & 0 \\
0 & 0 & 0 & 0 & \bar J & A_{(N+1)/2} & \ddots & 0 & 0 \\
0 & 0 & 0 & 0 & 0 & \ddots & \ddots & \bar J & 0 \\
0 & 0 & 0 & 0 & 0 & 0 & \bar J & A_3 & \bar J\\
\bar J & 0 & 0 & 0 & 0 & 0 & 0 & \bar J & A_2\\
\end{array}\right), \ \ \ 
A_n=1-\frac{ g^2}{\left(1+4g^2\bar{\alpha}_n^2\right)^{3/2}}{.}
\end{equation}
Let us consider a $(N-1)\times(N-1)$ submatrix of $\mathrm{Hess}_{N}$ that excludes the first column and first row, namely,
\begin{equation}\label{eq:subhessian}
\mathrm{Hess}'_{N}=
\left(\begin{array}{cccccccc}
A_2 & \bar J & 0 & 0 & 0 & 0 & 0 & 0 \\
\bar J & A_3 & \bar J & 0 & 0 & 0 & 0 & 0 \\
0 & \bar J & \ddots & \ddots & 0 & 0 & 0 & 0 \\
0 & 0 & \ddots & A_{(N+1)/2} & \bar J & 0 & 0 & 0 \\
0 & 0 & 0 & \bar J & A_{(N+1)/2} & \ddots & 0 & 0 \\
0 & 0 & 0 & 0 & \ddots & \ddots & \bar J & 0 \\
0 & 0 & 0 & 0 & 0 & \bar J & A_3 & \bar J\\
0 & 0 & 0 & 0 & 0 & 0 & \bar J & A_2\\
\end{array}\right),
\end{equation}
which respects the mirror symmetry about the minor-diagonal. Therefore, its eigenvector $\mathbf{x}$ should be an eigenvector of the corresponding symmetry operator. In order words, $\mathbf{x}'(j)=\pm \mathbf{x}'(N-j)$ for $j=1,2,\cdots,(N-1)/2$. If we choose an anti-symmetric eigenvector $\mathbf{x}'_{\mathrm{F}}$ that satisfies $\mathbf{x}'_{\mathrm{F}}(j)=-\mathbf{x}'_{\mathrm{F}}(N-j)$ with eigenvector $\lambda_{\mathrm{F}}$, then one can verify that $(0,\mathbf{x}'_{\mathrm{F}})^{\top}$ is an eigenvector of the original Hessian in Eq. (\ref{eq:symhessian}), namely,
\begin{equation}
	\mathrm{Hess}_{N}
	\left(\begin{array}{c}
		0\\\mathbf{x}'_{\mathrm{F}}
	\end{array}\right)=
	\left(\begin{array}{c}
		\bar J\left[\mathbf{x}'_{\mathrm{F}}(1)+\mathbf{x}'_{\mathrm{F}}(N-1)\right]\\\lambda_{\mathrm{F}}\mathbf{x}'_{\mathrm{F}}
	\end{array}\right)=
	\left(\begin{array}{c}
		0\\\lambda_{\mathrm{F}}\mathbf{x}'_{\mathrm{F}}
	\end{array}\right)=
	\lambda_{\mathrm{F}}
	\left(\begin{array}{c}
		0\\\mathbf{x}'_{\mathrm{F}}
	\end{array}\right)
\end{equation}
Therefore, in the frustrated superradiant phase of the Dicke lattice model for any odd $N$, there are always $(N-1)/2$ normal modes in the mean-field energy, which decouple a single unpaired site and anti-symmetric for the ferromagnetic pairs.
\end{proof}

\end{document}